\documentclass[aps,pre,twocolumn,superscriptaddress,showpacs,floatfix]{revtex4}
\usepackage{amssymb}
\usepackage{graphicx}
\usepackage{subfigure}	 	
\usepackage[english]{babel}
\usepackage{float}
\usepackage{subfigure}
\usepackage{color}

\begin{document}

\title{Peierls-Nabarro energy surfaces and directional mobility of discrete 
solitons in two-dimensional  saturable nonlinear Schr\"odinger lattices}

\author{Uta\ Naether}
\affiliation{Departamento de F\'isica, Facultad de Ciencias, Universidad 
de Chile, Santiago, Chile}
\affiliation{Center for Optics and Photonics, Universidad de Concepci\'on, 
Casilla 4016, Concepci\'on, Chile}

\author{Rodrigo A.\ Vicencio}
\affiliation{Departamento de F\'isica, Facultad de Ciencias, Universidad 
de Chile, Santiago, Chile}
\affiliation{Center for Optics and Photonics, Universidad de Concepci\'on, 
Casilla 4016, Concepci\'on, Chile}

\author{Magnus Johansson}
\email[]{mjn@ifm.liu.se}
\homepage[]{http://people.ifm.liu.se/majoh}
\affiliation{Department of Physics, Chemistry and Biology, Link\"oping 
University, SE-581 83 Link\"oping, Sweden}

\begin{abstract} 

We address the problem of directional mobility of discrete solitons in 
two-dimensional rectangular lattices, in the framework of a discrete 
nonlinear 
Schr\"odinger model with saturable on-site nonlinearity. A numerical 
constrained Newton-Raphson method is used to calculate two-dimensional 
Peierls-Nabarro energy surfaces, which describe a pseudopotential landscape 
for the slow 
mobility of coherent localized excitations, corresponding to continuous 
phase-space trajectories passing close to stationary modes. Investigating the 
two-parameter space of the model through independent variations of the 
nonlinearity constant and the power, 
we show how parameter regimes and directions of good mobility are connected 
to 
existence of smooth surfaces connecting the stationary states. In 
particular,  
directions where solutions can move with minimum radiation can be predicted 
from flatter parts of the surfaces. For such mobile solutions, 
slight perturbations in the transverse direction yield additional transverse 
oscillations with frequencies determined by the curvature of the energy 
surfaces, and with amplitudes that for certain velocities may grow rapidly. 
We 
also describe how the mobility properties and surface topologies are affected 
by inclusion of weak lattice anisotropy.

\end{abstract}

\pacs{42.65.Wi, 63.20.Pw, 63.20.Ry, 05.45.Yv}

\maketitle

\section{Introduction}
Intrinsically localized modes, or discrete solitons (breathers), appear as 
generic excitations in a large variety of physical systems \cite{rep2}, where 
spatial periodicity (or discreteness) provides gaps in the linear dispersion 
relation and nonlinearity allows for detuning the oscillation frequencies 
into 
these gaps, resulting in spatial localization. Particularly important 
applications 
of discrete nonlinear systems  are nonlinear optical waveguide arrays 
\cite{rep1},  with possibilities to change and control all essential 
parameters, such as geometry, dimensionality, nonlinearity, beam angle, etc. 
For weakly coupled waveguides, the relevant mathematical description, derived 
via coupled mode theory \cite{cj88,rep1}, is a Hamiltonian lattice equation 
of the discrete nonlinear 
Schr\"odinger (DNLS) type \cite{rep2,rep1,cj88,ej03}, where the particular 
type of nonlinearity 
depends on the nonlinear response of the medium.  The most studied case is a 
DNLS equation with cubic on-site nonlinearity\cite{ej03}, corresponding to a 
Kerr medium \cite{rep1}, which also appears generically as a modulational 
equation for the small-amplitude dynamics in chains of coupled anharmonic 
oscillators \cite{rep2}.   However, photorefractive media also enables the  
generation of discrete spatial solitons \cite{rep1}, and the corresponding 
lattice model is a DNLS 
equation with saturable nonlinearity (s-DNLS) which was studied in a number 
of theoretical works
\cite{prlkip,jpakhare,plaeil,prlmel,ob07,pre73,rnkf09}.

A particularly interesting property of the one-dimensional (1D) s-DNLS model 
\cite{prlkip,jpakhare,plaeil,prlmel,ob07}
is the existence of certain ``sliding velocities'', where  localized  discrete 
solitons may travel in the lattice without radiation. This behavior was 
connected to the existence of ``transparent points'' associated with the vanishing 
of a so-called Peierls-Nabarro (PN) potential barrier \cite{pre48}, usually 
defined as the difference in energy (Hamiltonian) at constant power (norm) 
between the two fundamental localized stationary solutions centered at one 
site (odd mode) and symmetrically in between two sites (even mode). Close to 
the points of vanishing of this energy difference are regions of stability 
exchange between the even and odd solutions, associated in general with the 
appearance of a family 
of intermediate, asymmetric stationary solutions, connecting both types of 
symmetric solutions at the bifurcation points \cite{pdaub}. The existence of 
regimes of enhanced mobility close to such bifurcation points is also 
well-known from studies of other one-dimensional lattice models 
\cite{pdaub,pdcret,oje03}. 

On the other hand, it is still an open issue whether moving discrete solitons 
may exist as localized, radiationless modes also in two-dimensional (2D) 
lattices. As was shown numerically in \cite{pre73}, a scenario with exchange 
of stability through bifurcations with asymmetric stationary solutions 
appears 
also for the 2D saturable model in a square (isotropic) lattice, involving in 
this case three different types of fundamental solutions \cite{krb00}: 
one-site (odd-odd, OO), two-site (odd-even, OE), and four-site 
(even-even, EE) modes. It was also shown numerically 
in \cite{pre73}, that solutions with good (but generally not radiationless) 
mobility in the axial directions may exist in these regimes, and that the 
necessary energy needed for rendering a mobile stable stationary solution 
agreed well with the concept of PN-barrier, if its definition was
extended to take into account also the energy for the relevant intermediate 
stationary solution 
(an analogous situation is well-known for the PN potential of 
kinks \cite{pr82}). Very similar observations were also made later for a 
2D DNLS model with cubic-quintic nonlinearity  \cite{ccmk09}. Moreover, in 
the 
low-power regime of the 2D s-DNLS equations, good mobility was also observed 
in diagonal directions \cite{pre73}.

However, the relation between existence of regions of stability exchange, 
small PN-barrier and mobile localized solutions in 2D is not that trivial, as 
shown, e.g., for a model with cubic inter-site nonlinearities in \cite{oj09}: 
even in regimes with small PN barrier and existence of intermediate 
solutions, 
the mobility may be very poor, if there is no continuous path in phase space 
passing close to the relevant stationary modes. On the other hand, as was 
observed for the low-power 
(i.e., close to continuum limit) regime of a 2D lattice with quadratic 
nonlinearity in \cite{susanto07}, the effective Peierls-Nabarro potential may 
in some situations be weak enough to allow mobility in {\em arbitrary} 
directions, without any direct connection to bifurcations and symmetry-broken 
stationary solutions.  

So there is clearly need for a better understanding of the conditions for 
mobility in 2D lattices. It is the purpose of the present paper to generalize 
the concept of PN-barrier as discussed above, and introduce a full 2D PN 
potential surface describing the  pseudo-potential landscape in-between all 
stationary modes. We will use a numerical constrained Newton-Raphson (NR) 
method, 
previously applied to 1D lattices in \cite{rodrigo1d}, to explicitly 
construct 
these surfaces for the 
2D saturable model from \cite{pre73}, and show how parameter regimes and 
directions of good 
mobility may be immediately identified from smooth, flat parts of these 
surfaces. We will also illustrate how the interplay between 
translational motion in one lattice direction and oscillatory motion in the 
orthogonal direction can be intuitively understood from the topology of the 
corresponding PN surfaces.

The structure of the paper is as follows. In Sec.\ \ref{model}, we describe 
the 
2D s-DNLS model, its stationary localized solutions (Sec.\ \ref{stat}), 
and the explicit implementation of the numerical constraint method used to 
calculate the PN potential surfaces (Sec.\ \ref{constraint}). 
We point out some essential differences between the 2D implementation and 
earlier implementations for the 1D case. In Sec.\ \ref{surfaces} we identify 
the regimes of parameter space where smooth surfaces can be found for the 
isotropic lattice, and make a classification of appearing surfaces with 
different topologies. In Sec.\ \ref{dynamics} we illustrate with direct 
dynamical simulations various types of behavior for discrete solitons moving 
slowly in different directions, corresponding to the surfaces discussed in 
Sec.\ \ref{surfaces}.  
We also here show examples on how simultaneous excitation of transverse 
oscillation modes may affect
discrete solitons moving slowly in axial directions.  
In Sec.\ \ref{anisotropy}, we analyze effects of adding a weak anisotropy to 
the lattices in the parameter regimes used in previous sections, in order 
mainly to investigate whether this may promote mobility in directions 
different from the axial or diagonal ones (which presumably would be 
associated with the appearance of additional "valleys" in the PN potential 
surface). Finally, in Sec.\ \ref{conclusions} we make some concluding 
remarks, 
summarizing our results and pointing out some relevant directions for future 
research.

\section{Model}\label{model}

We consider the following general form of the 2D s-DNLS equation in a 
rectangular geometry~\cite{pre73}:
\begin{equation}
i\frac{\partial U_{n,m}}{\partial z}+\Delta_{n,m}^\alpha U_{n,m}
-\frac{\gamma U_{n,m}}{1+|U_{n,m}|^2} = 0\ ,
\label{sdnls}
\end{equation}
where $z$ corresponds to the normalized propagation coordinate, 
$\gamma$ to an effective nonlinear parameter, and $U_{n,m}$ represents the 
(complex) electric field amplitude at site $\{n,m\}$ in an array of 
$N\times M$ sites. 
$\Delta_{n,m}^\alpha U_{n,m}\equiv (U_{n+1,m}+U_{n-1,m})+\alpha(U_{n,m+1}+U_{n,m-1})$ 
defines the linear interaction between nearest-neighbor waveguides, 
where $\alpha$ 
is an anisotropy parameter allowing us to study different 2D scenarios 
from a completely isotropic lattice ($\alpha=1$) to a completely decoupled 
set 
of 1D arrays ($\alpha=0$). Model (\ref{sdnls}) possesses two conserved 
quantities, the power (norm) defined as
\begin{equation}
P=\sum_{nm} |U_{n,m}|^2 \ ,
\label{pow}
\end{equation}
and the Hamiltonian (energy) defined as
\begin{eqnarray}
H=-\sum_{nm} \left[ \left(U_{n+1,m}+\alpha U_{n,m+1}\right)U_{n,m}^*
\right.\nonumber 
\\ \left. -\frac{\gamma}{2}\ln (1+|U_{n,m}|^2)+\text{c.c.}\right] .
\label{ham}
\end{eqnarray}
\subsection{Stationary solutions}
\label{stat}

Stationary solutions to Eq.\ (\ref{sdnls}) are of the form 
$U_{n,m}(z)=u_{n,m}\exp{(i \lambda z)}$, where $u_{n,m}$ is a $z$-independent 
(generally complex) amplitude and $\lambda$ corresponds to the propagation 
constant or 
frequency \cite{pre73}. 
Extended stationary solutions with constant amplitude (plane waves) 
exist in frequency bands, whose edges depend on the amplitude $|u_{n,m}|$.
The linear band region is obtained for low-amplitude plane waves 
($|u_{n,m}| \rightarrow 0$), and
it is easy to show that they will exist in the region 
$\lambda \in [-\gamma-2(1+\alpha),-\gamma+2(1+\alpha)]$.
Moreover, for high-amplitude plane waves ($|u_{n,m}| \rightarrow \infty$), 
the nonlinear term saturates and the corresponding interval is 
$[-2(1+\alpha),2(1+\alpha)]$.

\begin{figure}[htbp]
\centering
\includegraphics[width=0.48\textwidth]{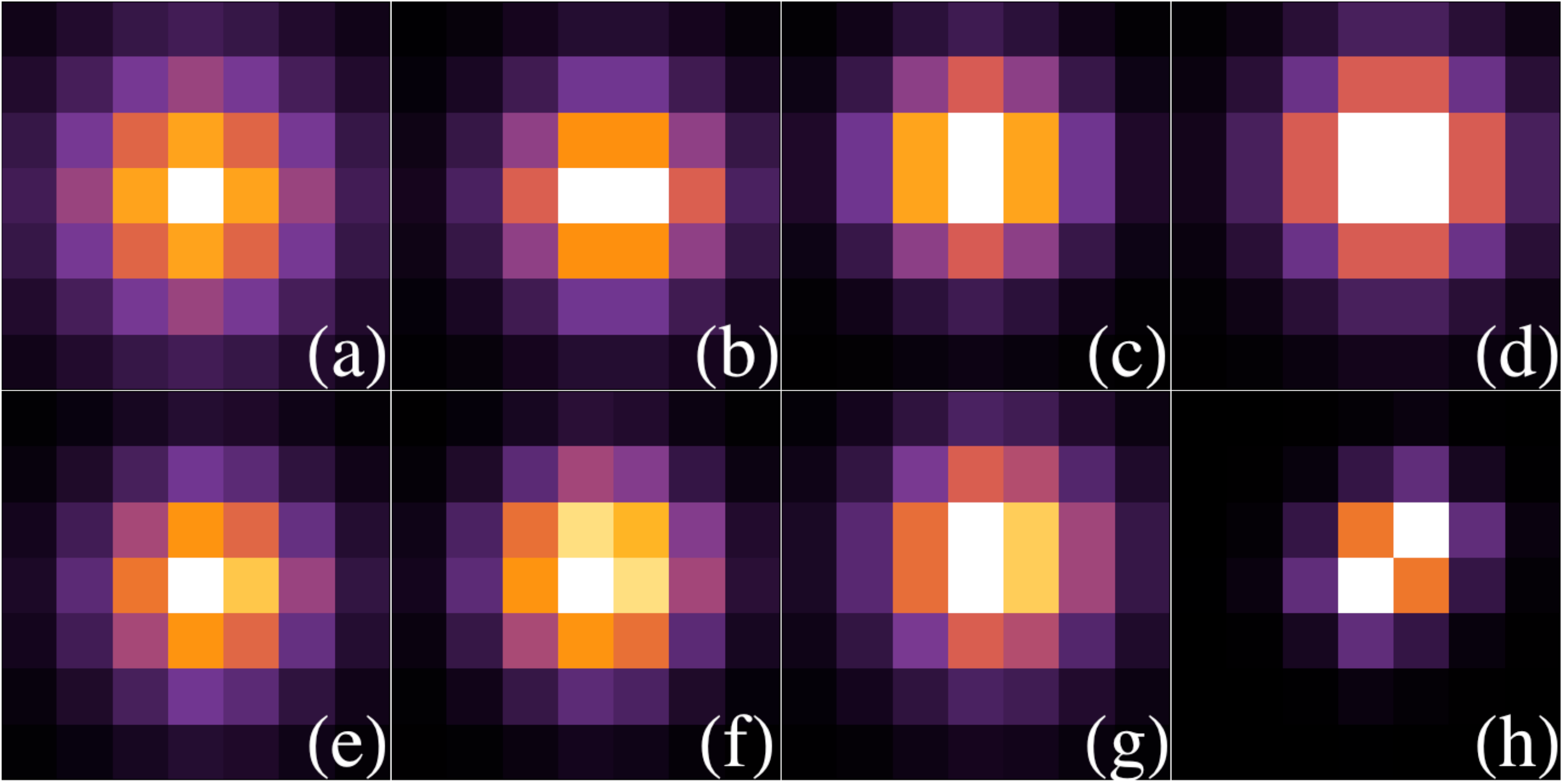}
\caption{(Color online) Examples of spatial profiles for (a) one-site (OO), (b) two-site horizontal (EO), (c) two-site vertical (OE), (d) four-site solutions (EE), (e) IS1, (f) IS2, (g) IS3-vertical, and (h) diagonal solutions, respectively, illustrated for an isotropic lattice ($\alpha = 1$).}
\label{profs}
\end{figure}

The fundamental localized stationary solutions to Eq.\ (\ref{sdnls}), 
defined as solutions with real amplitudes $u_{n,m}$ having a single 
maximum distributed on the sites in one unit-cell of the lattice, were 
described for the isotropic case in \cite{pre73}. In general, we may 
identify one-site (OO), two-site horizontal (EO), two-site vertical (OE), 
and four-site (EE) solutions.  
Figs.\ \ref{profs}(a)-(d) show typical profiles for these types of excitations.
As discussed in \cite{pre73}, 
these in-phase localized solutions bifurcate from the fundamental linear mode
(upper band edge) in the small-amplitude limit, and merge into the 
corresponding high-amplitude plane-wave mode in the infinite-power limit.
As a consequence, they exist in the region 
$\lambda\in [-\gamma+2(1+\alpha),2(1+\alpha)]$, so that the size of the 
existence region will be just $\gamma$. 
[A two-site diagonal solution \cite{Kalosakas} [see Fig.\ \ref{profs}(h)] could also be excited but only 
in frequency regimes where solutions are very localized with
$\lambda \lesssim 0$, requiring large values of $\gamma $ 
(for $\alpha=1$, $\gamma \gtrsim 8.2$)].
In addition, in certain parameter regimes there are also asymmetric, 
intermediate stationary solutions, associated with exchange of stability 
between the symmetric ones \cite{pre73}.
We may identify three types of such solutions, termed henceforth
intermediate 1 (IS1), intermediate 2 (IS2), 
and intermediate 3 (IS3), respectively. IS1 and IS3 both connect 
a stable two-site mode, OE/EO, with another simultaneously stable solution: 
the one-site mode OO (IS1), or the four-site mode EE (IS3) [see Figs.\ \ref{profs}(e) and (g)].
Thus, in these cases the unstable intermediate solutions act as carriers of instability between the 
corresponding fundamental modes. The unstable IS2 solution exists when the two-site solutions are stable. It connects the unstable one-site solution OO with the likewise unstable four-site solution EE, and stabilizes the latter mode when reaching it [see Fig.\ \ref{profs}(f)].

\begin{figure}[htbp]
\centering
\includegraphics[width=0.46\textwidth]{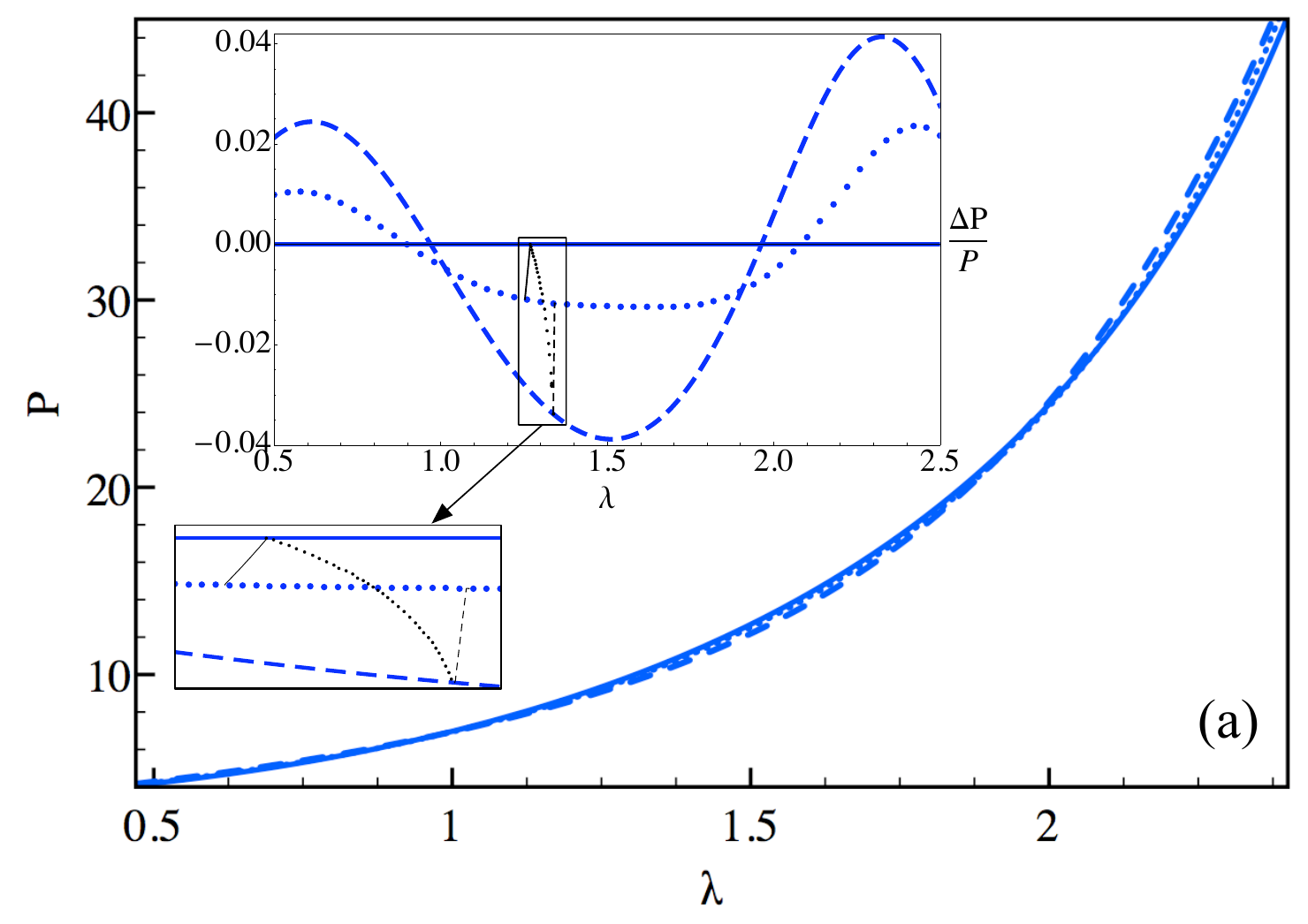}\vspace{0.1cm}
\includegraphics[width=0.46\textwidth]{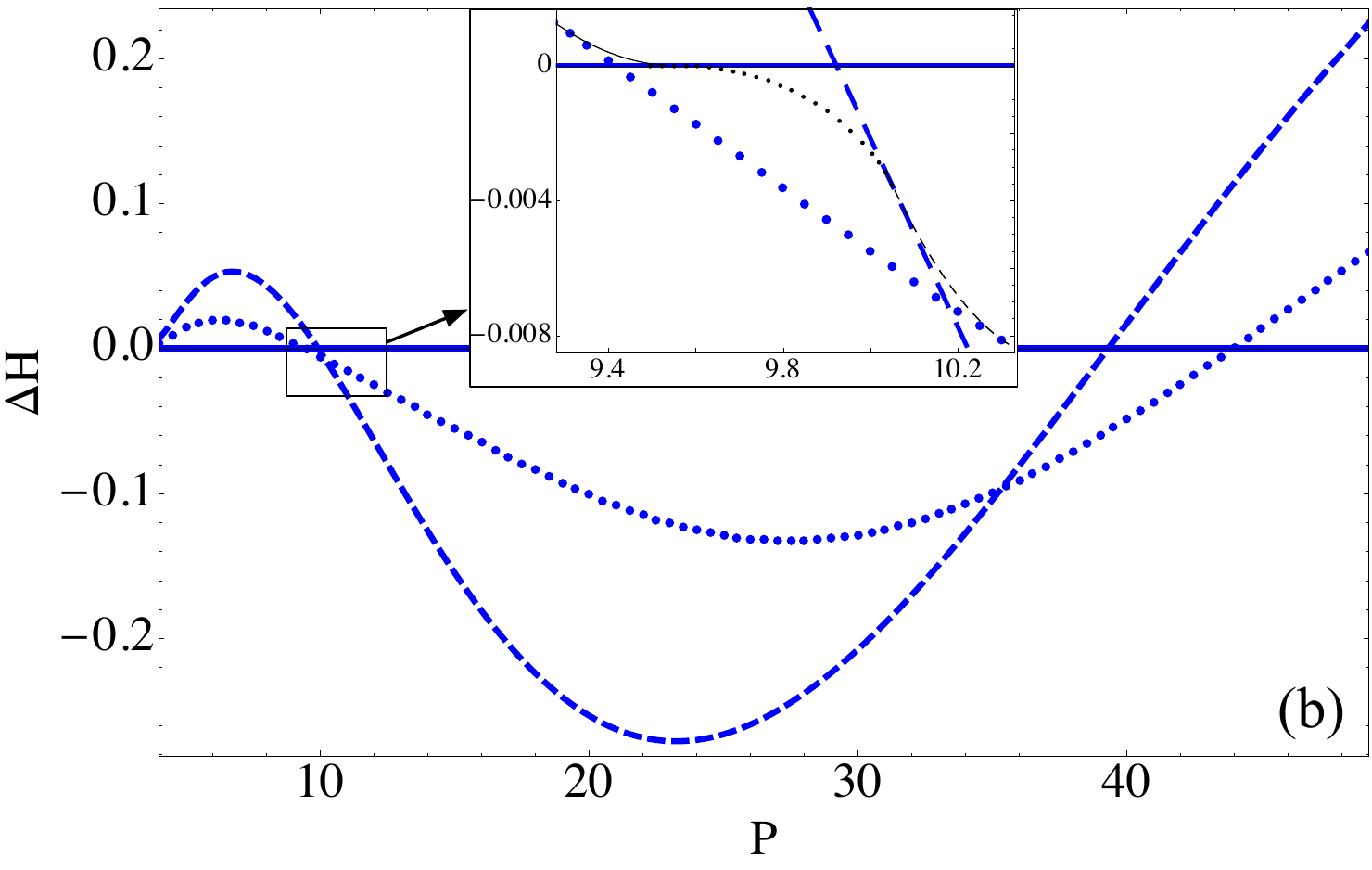}\vspace{0.05cm}
\includegraphics[width=0.46\textwidth]{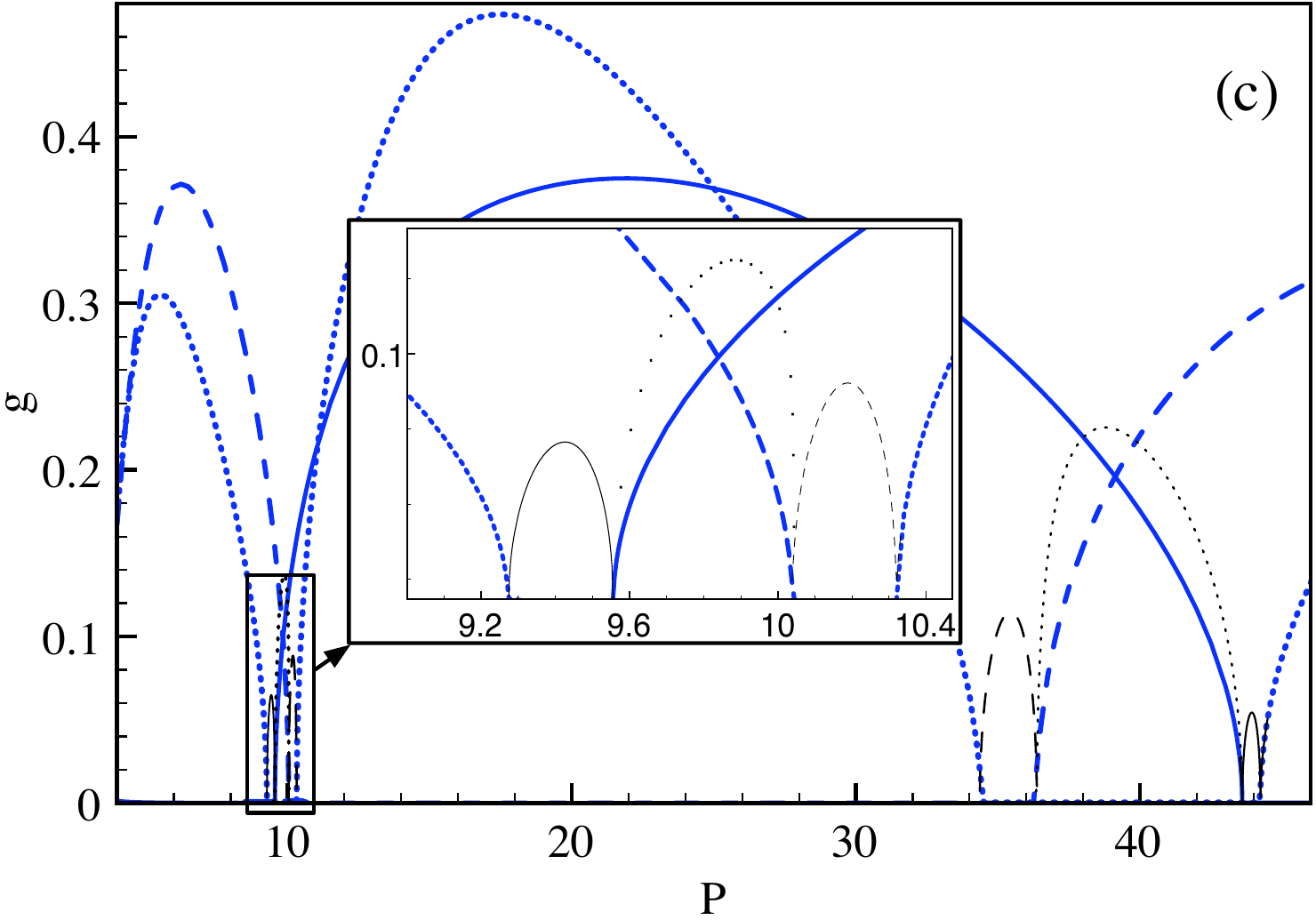}
\caption{(Color online) Properties of fundamental stationary solutions for 
$\gamma=4$ and $\alpha=1$: 
(a) Power versus frequency; (a)-inset: $\Delta P/P$ versus $\lambda$; (b) $\Delta H$ versus power; (c) Stability versus power.
Different solutions are plotted with different line-types: 
one-site (thick solid blue), two-site (thick dotted blue), 
four-site (thick dashed blue), IS1 (thin solid black), 
IS2 (thin dotted black), and IS3 (thin dashed black), respectively.
}\label{PL}
\end{figure}

Figs.~\ref{PL} shows fundamental properties of different stationary solutions
for an isotropic lattice with $\gamma = 4$ (similar curves were plotted and 
discussed in 
\cite{pre73}, but we here include some additional examples to facilitate 
the comparison with the full PN surfaces in Sec.\ \ref{surfaces}). Power versus frequency curve is shown in Fig.~\ref{PL} (a), where different solutions repeatedly cross each other in 
the whole range of parameters. In Fig.~\ref{PL} (a)-inset we plot $\Delta P/P\equiv (P_i-P_{OO})/P_{OO}$ as a function of $\lambda$, for $i=OO, EO/OE, EE, IS1, IS2, IS3$. This figure shows clearer how solutions cross each other including many intermediate solutions appearing in the region $\lambda\sim 1.3$. 
Moreover, we also observe these crossings when plotting $\Delta H\equiv H_i-H_{OO}$ versus power [Fig.~\ref{PL} (b)]. (We plot $\Delta H$ instead of $H$ because Hamiltonian differences between stationary solutions of the same power are generally quite small, which is indeed a
favorable scenario for moving solutions).
Inset in Fig.~\ref{PL} (b) shows a zoom in the region $P\sim 9.8$ where the 
Hamiltonian of different fundamental solutions matches at some points.
This was initially interpreted as a vanishing PN barrier~\cite{prlkip}, 
which is actually not the case due to the intermediate solutions appearing 
in-between, raising the effective energy barrier~\cite{pre73} [see Fig.~\ref{PL} (b)-inset when thick blue lines coincide]. However, good 
mobility would be expected close to these regions but, still, it will be 
strongly determined by the specific kick or perturbation given to the 
solution 
in order to put it in movement.

In order to study the stability of stationary localized solutions we 
implement 
a standard method \cite{jpakhare,pre73}, obtaining all the stable/unstable 
linear 
perturbation-modes and their corresponding instability measure 
$g=\max_j[\Im \{\omega_j\}]$, 
where $\omega_j$ are the oscillation frequencies 
of the linear eigenmodes.  Thus, if $g=0$ 
the solution is (at least marginally) stable and if $g>0$ it is unstable. 
Stability versus power for the same solutions as in Figs.\ \ref{PL} (a), (b)
is 
shown in Fig.\ \ref{PL} (c). (Note that in the corresponding figures 2-3 of 
\cite{pre73}, the quantity $G=-g^2$ was used as instability measure.) 
From this figure, we can see how unstable 
intermediate solutions IS1, IS2, IS3 
appear when two or three solutions 
(regarding OE and EO as different solutions)
are simultaneously stable. 
Note that we never observed regions for simultaneously 
stable one and four-site solutions for  isotropic coupling. However, 
as will be discussed further in Sec.\ \ref{anisotropy}, in the anisotropic 
case 
such regions exist.

\subsection{Constraint method}
\label{constraint}

The constraint method allows us to construct energy surfaces connecting 
stationary solutions for a given value of power. In that sense, it helps us 
to effectively predict and interpret 
the dynamics across the lattice. Critical points will 
represent stationary solutions, and a coherent 
movement across the lattice should transform one solution to the other by 
keeping the power constant~\cite{pre48}. 
The method was originally introduced by Aubry and 
Cretegny~\cite{pdaub} and then implemented by Savin et al.~\cite{savin} to study the mobility of kinks in nonlinear Klein-Gordon lattices (see also Ref.\ \cite{Sepulchre} 
for a related approach to analyze travelling 
breathers in 1D oscillator chains in terms of an effective Hamiltonian.) 
Lately, it was numerically 
implemented to analyze surface states 
in one-dimensional semi-infinite systems~\cite{rodrigo1d}.
By adiabatically changing the amplitude in 
one 
particular site, specifically chosen as the one after the main peak 
(the peak is at $n_c$ and the constrained amplitude at $n_c+1$), 
the one- and the two-site solutions could be connected. Ending the sweep 
when $u_{n_c}=u_{n_c+1}$ and the center of mass of the constrained solution, 
$X$, has varied from $n_c$ to $n_{c}+0.5$, 
a one-dimensional energy surface, $H$ vs $X$, can be sketched. 
Technically speaking, the method used in Ref.\ \cite{rodrigo1d} consists on 
eliminating one equation from the Newton-Raphson problem, the one of the 
constrained amplitude which is not anymore an unknown variable. However, as 
the power is kept constant, an equation for $P$ is added and the 
frequency $\lambda$ becomes a variable completing the variable-equations set. 

In the present work, for the 2D lattice we implement  
a more sophisticated method 
where we explicitly vary the center of mass instead of the amplitude.
Since we are interested in the energy landscape around the fundamental 
stationary solutions, we will assume also the amplitude  $u_{n,m}$ of the
constrained solutions to be real 
and positive on the constraint sites. 
Then, from the definition of the center of mass coordinates 
in the horizontal and vertical directions,
\begin{equation}
X\equiv\frac{\sum_{nm} n |u_{n,m}|^2}{P}\ \ \ \ \ \ 
\text{and}
\ \ \ \ \ \ Y\equiv\frac{\sum_{nm} m |u_{n,m}|^2}{P}\ , 
\label{xy}
\end{equation}
we can solve for any pair of amplitudes $A \equiv  u_{n_A,m_A} $ and 
$B\equiv  u_{n_B,m_B } $ at places $\{n_A,m_A\}$ and $\{n_B,m_B\}$:
\begin{eqnarray}
A&=&\sqrt{\frac{m_B S_n-n_B S_m +P(n_B Y-m_BX)}{n_Bm_A-n_Am_B}}\nonumber\\
&&\label{cmab}\\
B&=&\sqrt{\frac{n_A S_m -m_A S_n +P(m_A X-n_A Y)}{n_Bm_A-n_Am_B}}\nonumber
\label{cma}
\end{eqnarray}
where $S_j\equiv\sum_{nm}(1-\delta_{n,n_A}\delta_{m,m_A}-
\delta_{n,n_B}\delta_{m,m_B}) j  |u_{n,m}|^2$ with $j=n,m$. 
Thus, as before the actual 
constraints will be in the amplitudes $A$ and $B$, but now we can tune the 
center of mass 
as wished, from a given 
stationary solution towards any other. 
In general, the constrained solutions obtained in this way will of course  
not be stationary solutions of the full system. To
identify any stationary solution - including the intermediate ones -  we 
check whether the value of $\lambda$ obtained from the constrained 
NR scheme coincides with the frequency of a hypothetical 
true stationary solution to 
the full  Eq.\ (\ref{sdnls}) with the computed amplitude profile. 
Furthermore, constrained solutions allow us 
to calculate their Hamiltonian, and therefore to construct an effective 
energy landscape. 

\begin{figure}[htbp]
\centering
\includegraphics[width=0.45\textwidth]{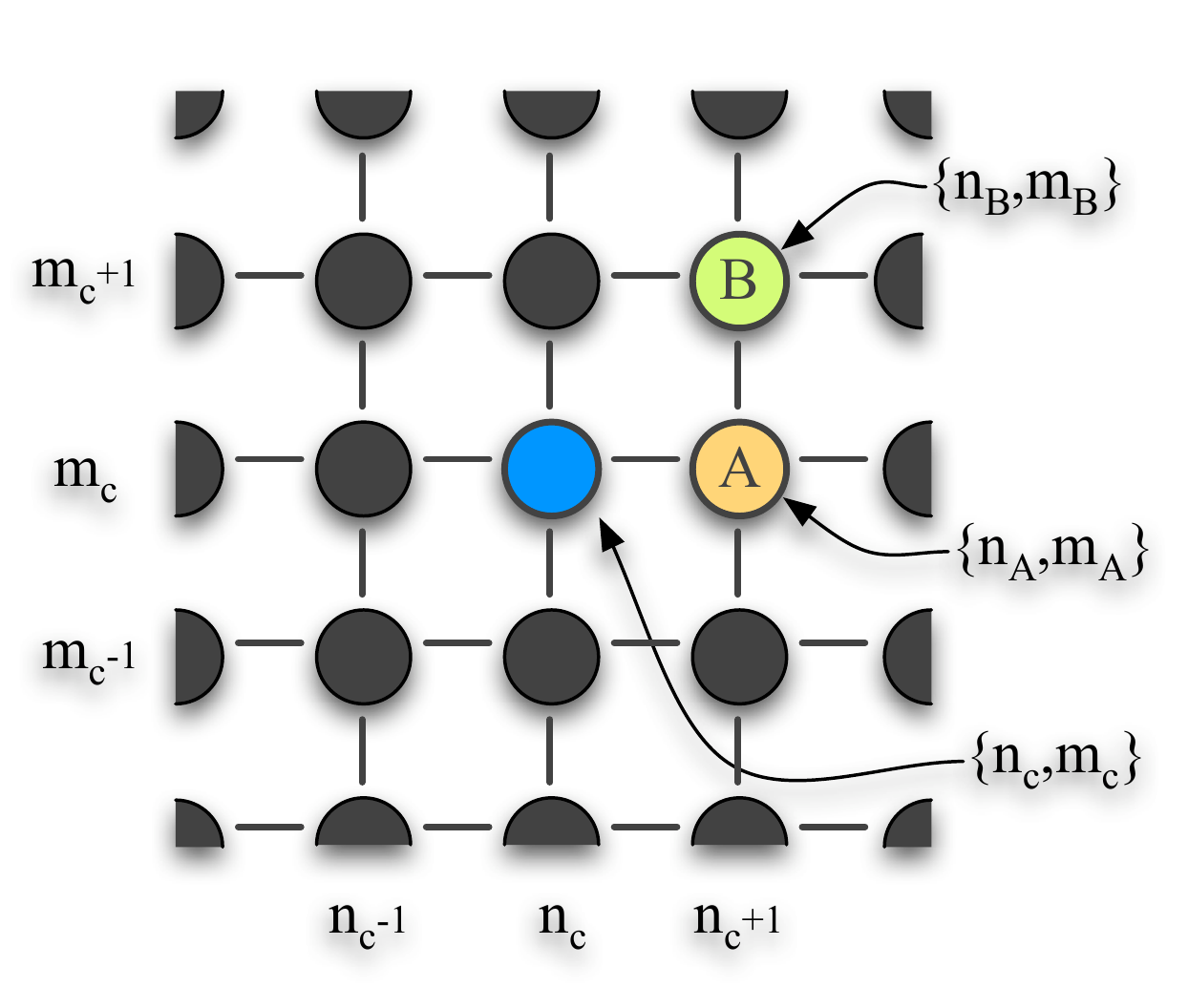}
\caption{(Color online) 
Array scheme showing the constraint locations used to obtain 
the energy surfaces below. }
\label{scheme}
\end{figure}
%
The choice for positions $\{n_A,m_A\}$ and $\{n_B,m_B\}$ is evidently not 
unique, but in order to most efficiently trace out a smooth energy landscape 
connecting the fundamental solutions (if it exists), the constraint 
sites should preferably be chosen within the unit cell where the 
amplitudes are large.  
It turned out that, starting from a stationary 
one-site solution centered at 
$\{n_c,m_c\}$, in most cases 
the best option is to choose the first constraint at 
site $\{n_c+1,m_c\}$ or $\{n_{c},m_c+1\}$, and the other one at site 
$\{n_c+1,m_c+1\}$, as sketched in Fig.\ \ref{scheme}. The results shown 
in the following sections are obtained using these constraint sites. We 
also tried using constraint sites at $\{n_c+1,m_c\}$  and $\{n_{c},m_c+1\}$
(i.e., along a diagonal); however, with this choice 
we typically were not able to find the four-site EE solution when starting 
from the one-site OO. Instead, the NR method with constraints on these sites 
generally converges, 
at $X=n_c+\frac{1}{2}, Y=m_c+\frac{1}{2}$, to a two-site diagonal solution 
(which as remarked in Sec.\ \ref{stat} is an exact solution to the 
unconstrained equation only for large $\gamma$), which is not expected to 
be relevant for the mobility process. 

\section{Isotropic lattice}

\subsection{Description of PN surfaces}\label{surfaces}

\begin{figure*}[h]
\includegraphics[width=0.45\textwidth]{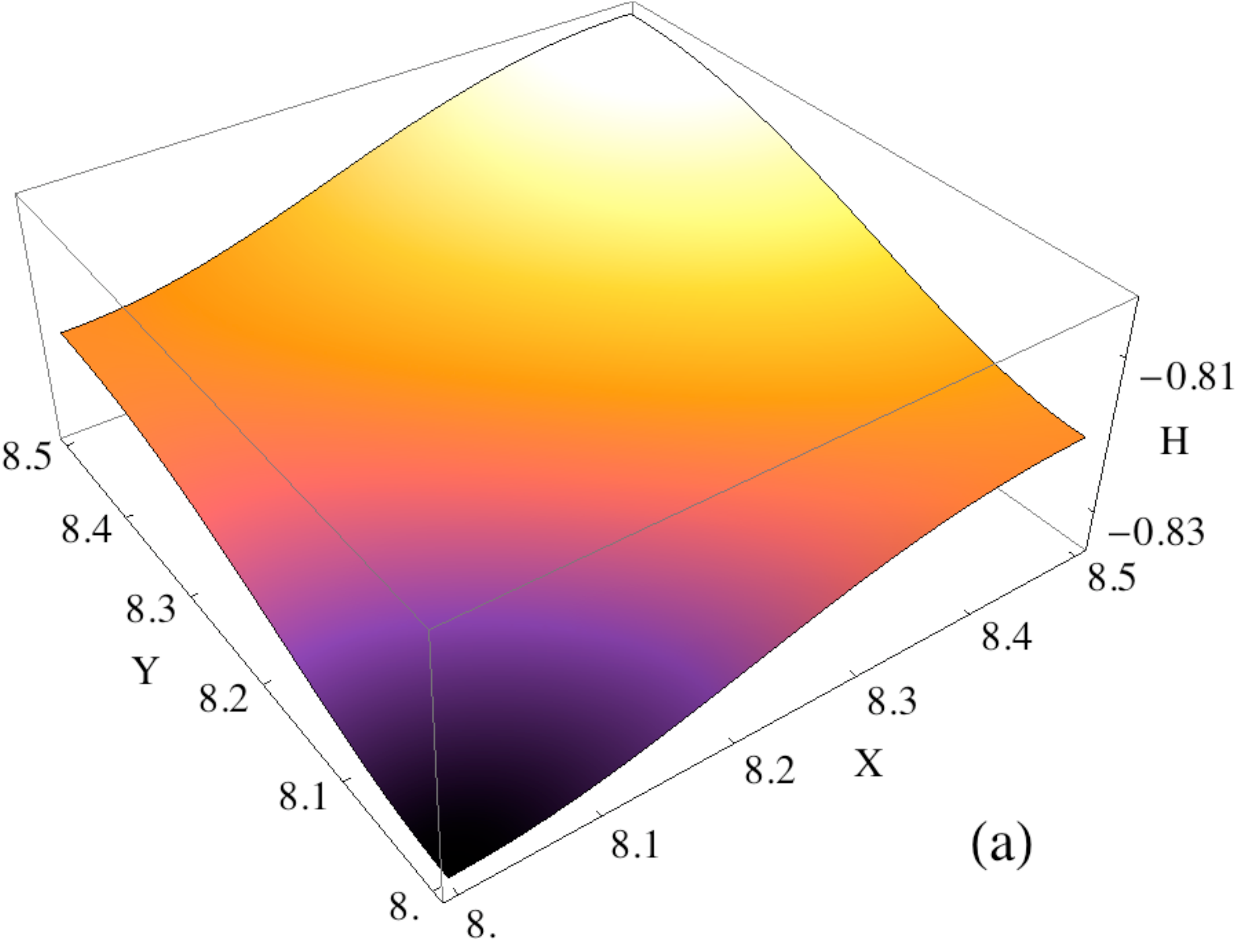}
\includegraphics[width=0.45\textwidth]{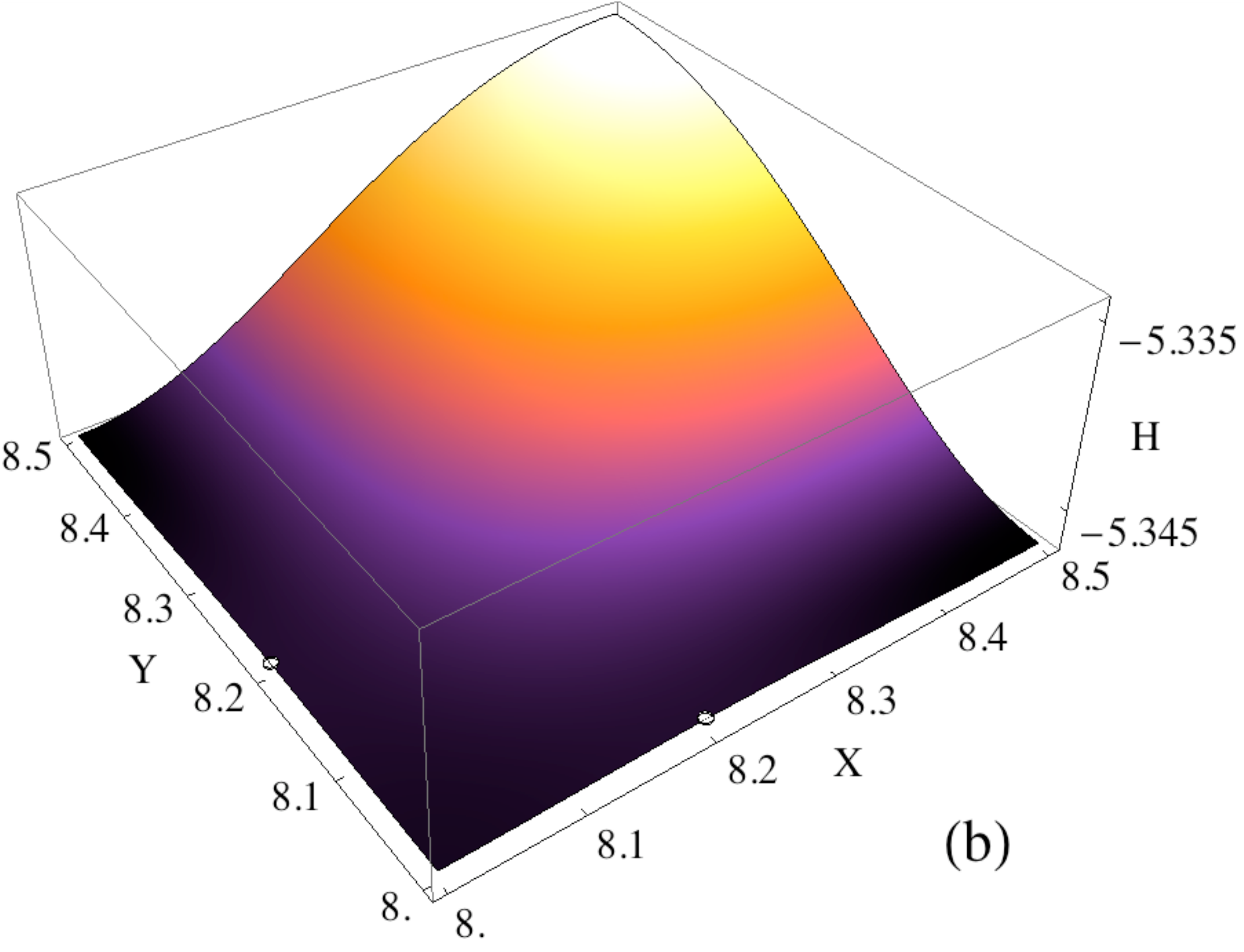}
\includegraphics[width=0.45\textwidth]{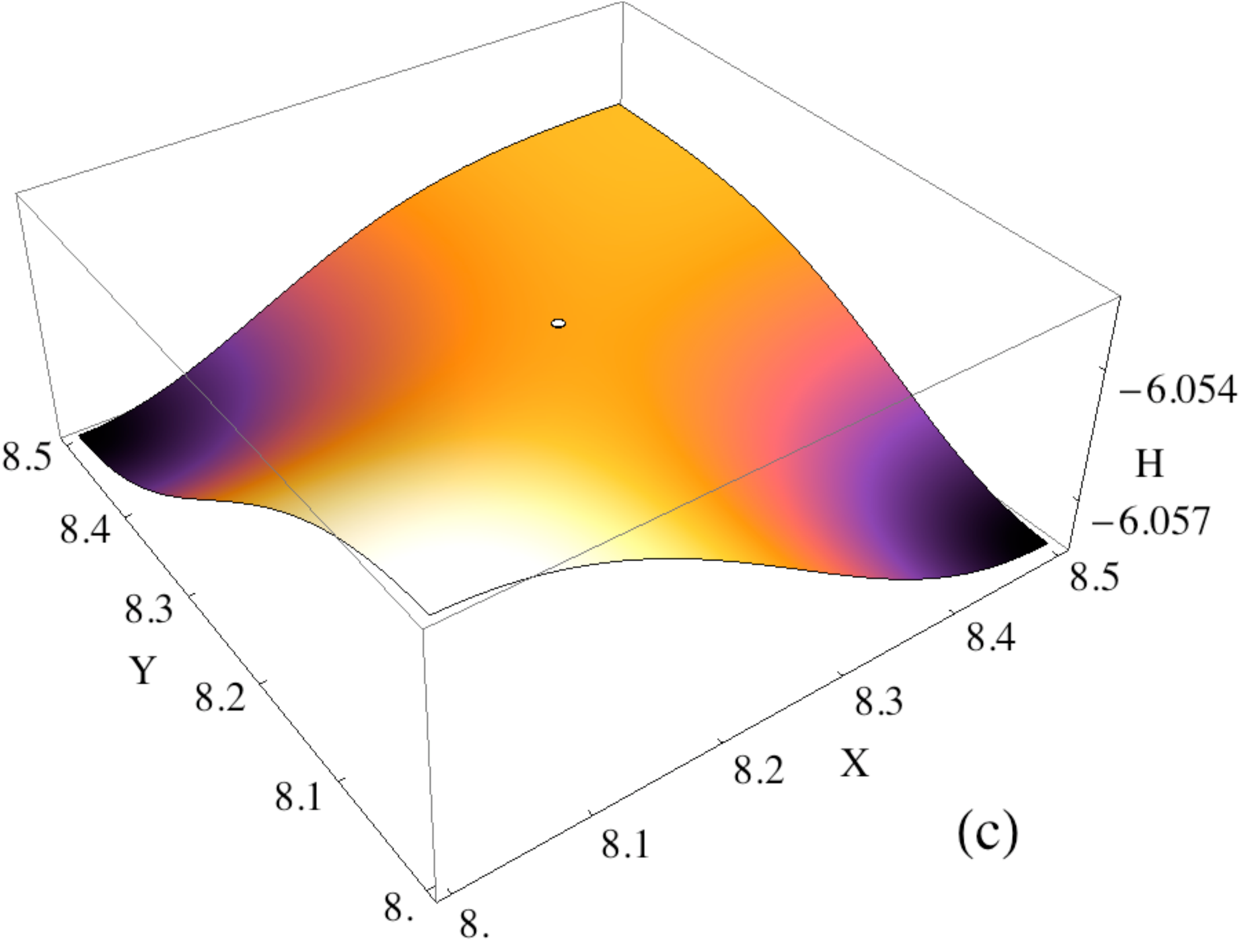}
\includegraphics[width=0.45\textwidth]{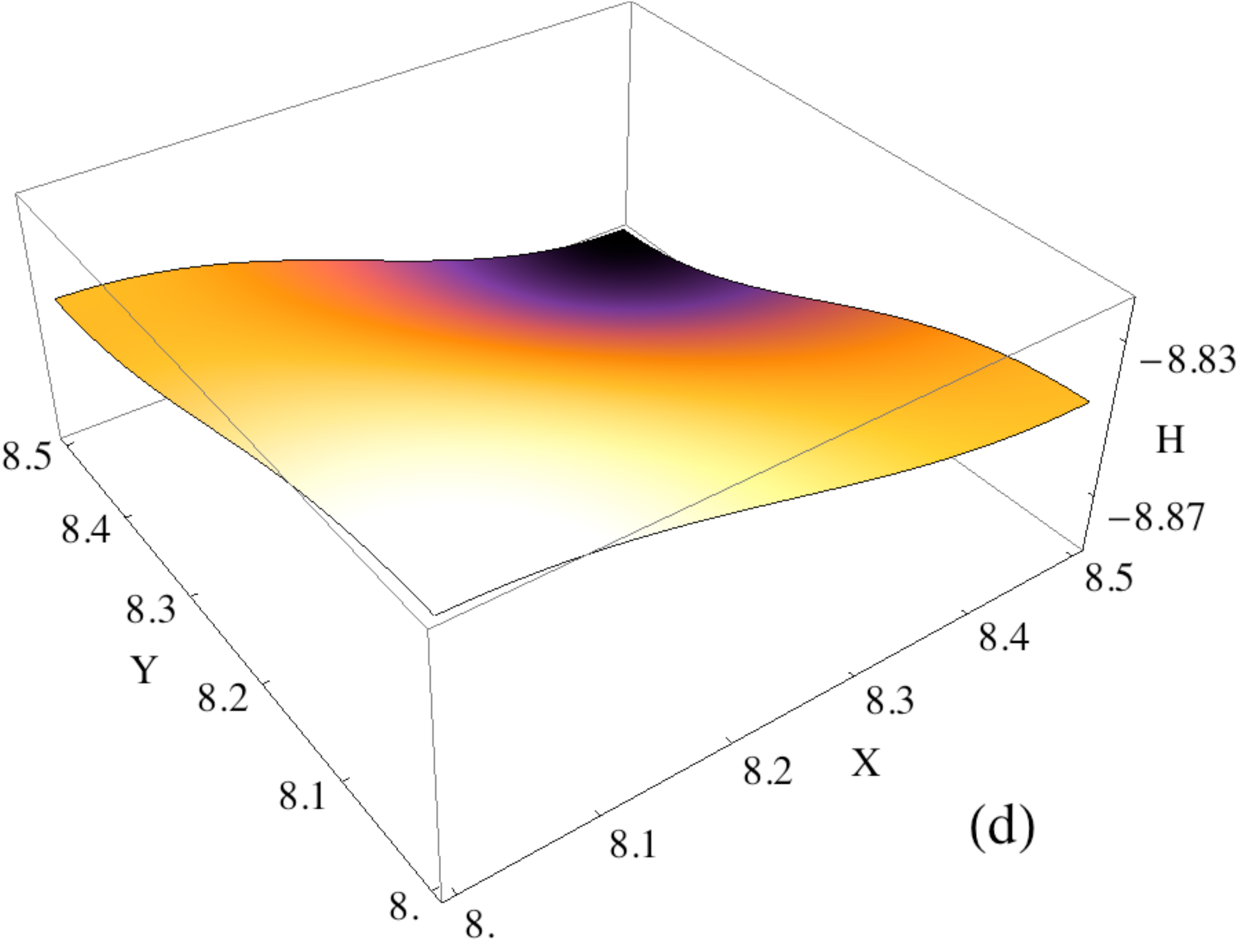}
\includegraphics[width=0.45\textwidth]{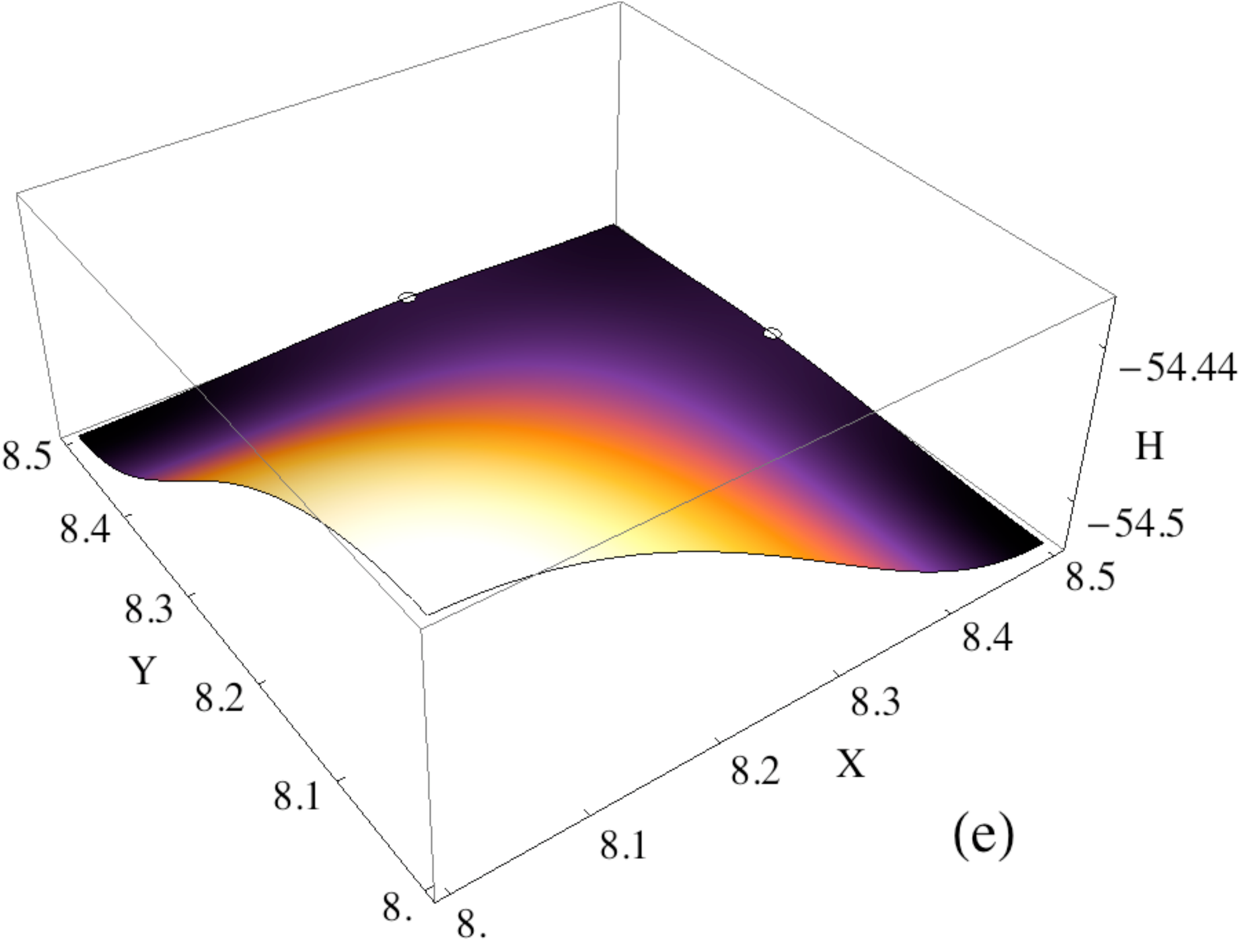}
\caption{(Color online) 
Energy surfaces for $\gamma=4$, $\alpha=1$, in the five different 
power regimes discussed in the text : (a) $P=5$; (b) $P=9.45$; (c) $P=10$; 
(d) $P=12$; (e) $P=35.5$. The center of mass $\{X,Y\}$ for the four stationary 
solutions are: 
$\{8,8\}$ (OO), $\{8.5,8\}$ (EO), $\{8,8.5\}$ (OE), and $\{8.5,8.5\}$ (EE).
White dots denote local extrema corresponding to intermediate solutions.
(System size $N=M=15$ and fixed boundary conditions were used.)
}
\label{pot1}
\end{figure*}

By exploring the parameter-space $(\gamma, P)$, we have identified regions 
where we were able to 
compute complete two-dimensional energy surfaces (using constraint as shown in 
Fig.\ \ref{scheme}) and regions where we cannot. In general, 
complete surfaces could not be obtained for large values of 
the nonlinearity constant, $\gamma \gtrsim 6$, when solutions get strongly 
localized. In some cases we could trace some specific mobility directions
(e.g., along lattice axes or  diagonal) connecting stationary solutions 
with almost equal Hamiltonian, but not 
the full two-dimensional scenario involving all fundamental solutions.
Note also that, as remarked in \cite{pre73}, for large $\gamma$ and small 
power the s-DNLS model essentially behaves as a cubic DNLS model with 
effective nonlinearity $\gamma P$.
Consequently, we could not find complete
energy surfaces with all fundamental solutions for the cubic DNLS either
(although surfaces involving the two-site diagonal solution in place of the 
EE solution could be obtained as mentioned in Sec.\ \ref{constraint}; 
for the cubic DNLS this two-site 
solution does exist as a true stationary solution above some threshold 
level of power \cite{Kalosakas}).

On the other hand, for very small values of $\gamma$, the computation of 
energy surfaces becomes difficult for technical reasons: due to the widening 
of the solutions, considerably larger lattices are needed to remove the 
influence from boundary effects. Therefore, in
the following, we will present the main phenomenology for the isotropic 
case ($\alpha=1$) found for intermediate values of $\gamma$,  
$3 \lesssim \gamma \lesssim 5$, where complete 
surfaces involving all four fundamental solutions 
are obtained for all values of the power. We will show 
results for the particular value $\gamma=4$, but 
the scenario is found to be qualitatively the same for all $\gamma$ in this 
interval.
 
From Fig.\ \ref{PL} (c), we may identify essentially
five different regimes where energy surfaces of qualitatively different 
nature 
should be expected, depending on the level of power. 
The first 
one is for low power, where (similarly to the cubic DNLS) 
the one-site solution is always stable
and the other fundamental stationary solutions are all  
unstable. The 
corresponding energy surface is illustrated in  Fig.\ \ref{pot1} (a), where 
the one-site solution yields the energy minimum, the two-site solutions 
saddle points and the four-site solution the maximum. Note that, in this 
low-power regime, the surfaces for this value of $\gamma$  are 
still rather flat, and therefore some mobility may result if the one-site 
solution is kicked to overcome the barriers, in the axial as well as in 
the diagonal directions, as illustrated by Figs.\ 4 (a) and (c) 
in Ref.\ \cite{pre73}. (Another example of mobility in this regime is discussed
in Sec.\ \ref{dynamics} below.)

The saturable nature of the system becomes evident at higher powers. In the 
second regime, appearing for the first time when 
$9.27 \lesssim P \lesssim 9.55$,  the one- and the two-site 
solutions are stable simultaneously  
and, as shown in  
Fig.\ \ref{pot1} (b), these three points all correspond to local 
minima of the surface. Intermediate solutions IS1 connecting the one- with the 
two-site solutions in the 
horizontal and vertical directions appear 
as saddle points [white dots in  
Fig.\ \ref{pot1} (b)]. 
Note that the energy landscape for the parameter values 
in Fig.\ \ref{pot1} (b) is almost flat between the one-site and 
two-site solutions in the axial directions, 
leading to the very good axial mobility shown  in Fig.\ 4 (d) 
of Ref.\ \cite{pre73}, while the maximum corresponding to the 
unstable four-site solution creates a 
too large effective barrier to overcome in the diagonal direction.

The third  power region is the one in which only the two-site solutions 
are stable. 
It  appears for the first time when $9.55 \lesssim P \lesssim 10.04$, 
and is illustrated 
in  Fig.\ \ref{pot1} (c). Here, the stable two-site solutions 
correspond to two local minima of the surface, and the unstable one- and 
four-site solutions both to local maxima. The two unstable solutions are 
connected by the intermediate solution IS2 (white dot at the surface), 
corresponding to a saddle point which for symmetry reasons 
(for the isotropic lattice) will lie 
along the diagonal connecting the unstable solutions. 
As will be illustrated in 
Sec.\ \ref{dynamics}, the easiest mobility  in this case is expected to occur 
in a diagonal direction, connecting the two stable stationary solutions.

The fourth regime corresponds to simultaneously stable two- and four-site 
solutions.  Such a regime appears for the first 
time when 
$10.04 \lesssim P \lesssim 10.32$. 
If we further 
increase the power, we will find a similar region with these three 
simultaneously stable solutions for 
$34.5\lesssim P \lesssim 36.25$. In this regime, as illustrated in  
Fig.\ \ref{pot1} (e), 
the two- and four-site solutions all correspond to minima, 
the unstable
intermediate solutions IS3 to saddle points,  and 
the one-site solution to a maximum. 
Thus, comparing Figs.\ \ref{pot1} (b) and (e), 
we see that the structure of the potential has been completely inverted. 
Now, a big ``hill'' is located at the one-site 
solution, and as a consequence no 
mobility is expected involving this solution. As will 
be shown below in  Sec.\ \ref{dynamics}, the simplest mobility 
scenario will now be the one in which the two-site solutions travel through 
the 
four-site one across the lattice.

The fifth regime, with the four-site solution being the only stable one, 
occurs for the first time when 
$10.32 \lesssim P \lesssim 34.5$.
As illustrated in Fig.\ \ref{pot1} (d),  the four-site solution 
is now a  minimum of the PN potential,  
while the other three unstable 
solutions correspond to saddle points (EO and OE) and maximum (OO), 
respectively. (There are no intermediate solutions in this regime.) 
Thus, by 
increasing the power  
we have now
reverted the
surface compared to the low-power regime in Fig.\ \ref{pot1} (a). 

Further increasing of power shows for $36.25 \lesssim P \lesssim 43.6$ a 
new region of stable two-site solutions, corresponding to the third regime. 
For $43.6 \lesssim P \lesssim44.2$  the one- and two-site solutions are 
simultaneously stable, so the scenario is equivalent to the second regime, 
followed again for  $P\gtrsim 44.2$ by the first regime, 
respectively. This second 
complete inversion thus corresponds to a regain of the low-power 
characteristics  of the surface. 
Furthermore, repetition of these scenarios can be found for $P\gtrsim 118$, 
and we confirmed, e.g., the existence of a complete, smooth surface, analogous 
to the one shown in Fig.\ \ref{pot1} (d) for the fifth regime, for $P=120$.

\subsection{Mobility dynamics for isotropic lattice}\label{dynamics}

To explicitly show the connection between the different types of 
energy surfaces described in Sec.\ \ref{surfaces} and the mobility of 
localized solutions, 
we numerically integrate model\ (\ref{sdnls}) by taking as initial condition a
stationary solution perturbed with a small
kick: $U_{n,m}(0)=u_{n,m} \exp[i k_x(n-n_c)+i k_y(m-m_c)]$, 
where $k_x$ and $k_y$ correspond to the kick strength 
in the horizontal and vertical 
directions, respectively. 
If the surfaces would be completely flat, 
solutions should move even with an infinitesimally small kick. 
However, as we have seen, in general surfaces are not flat due to  
the discreteness and the self-induced PN potential, and although the PN
barrier can be very small in certain directions, it is generally non-zero.  
Therefore, in order to put a localized solution in movement, 
some amount of kinetic energy (represented by this kick) should be given to 
effectively overcome the energy barriers.

\begin{figure*}[htbp]
\includegraphics[width=0.45\textwidth]{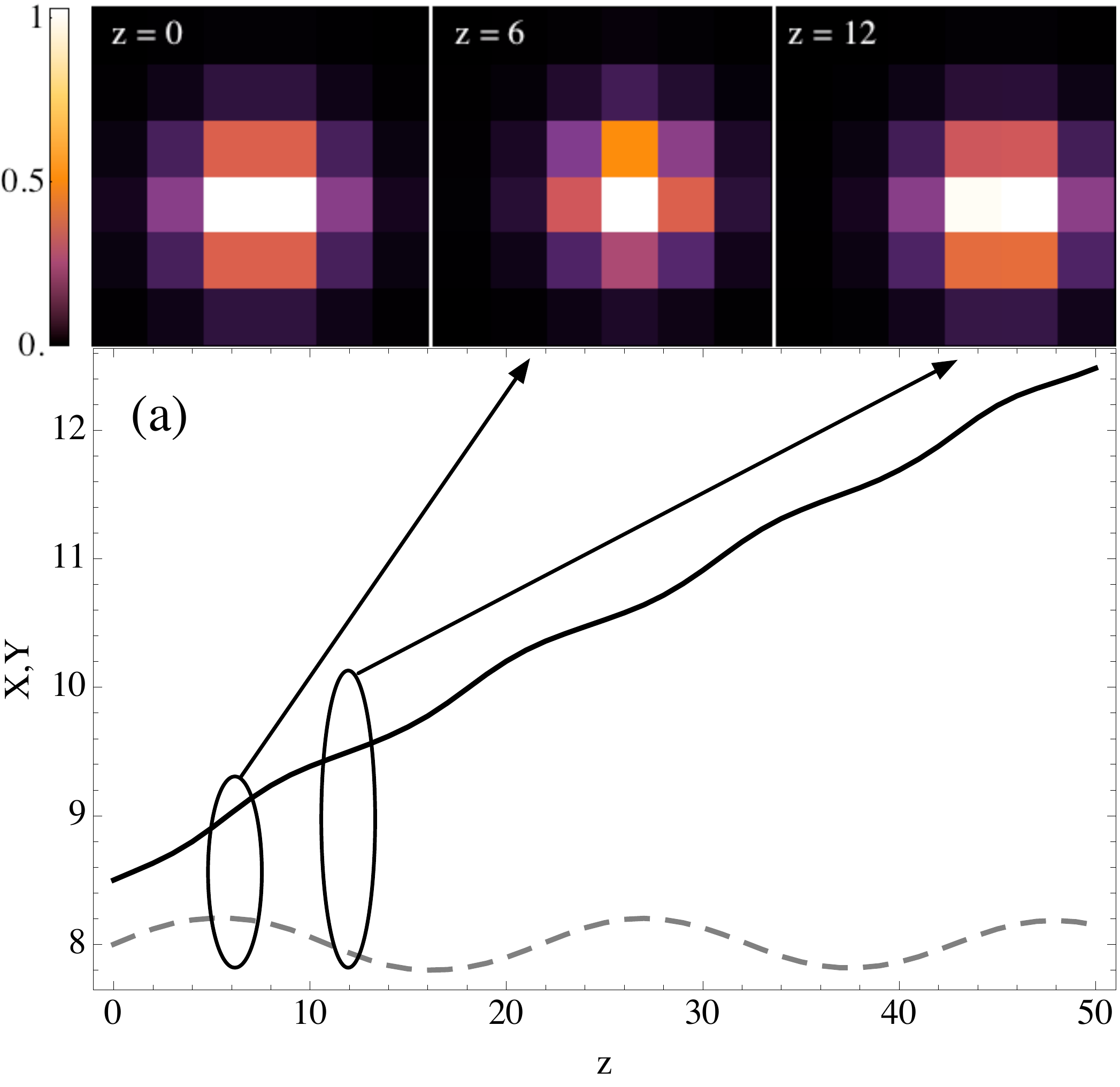}
\includegraphics[width=0.45\textwidth]{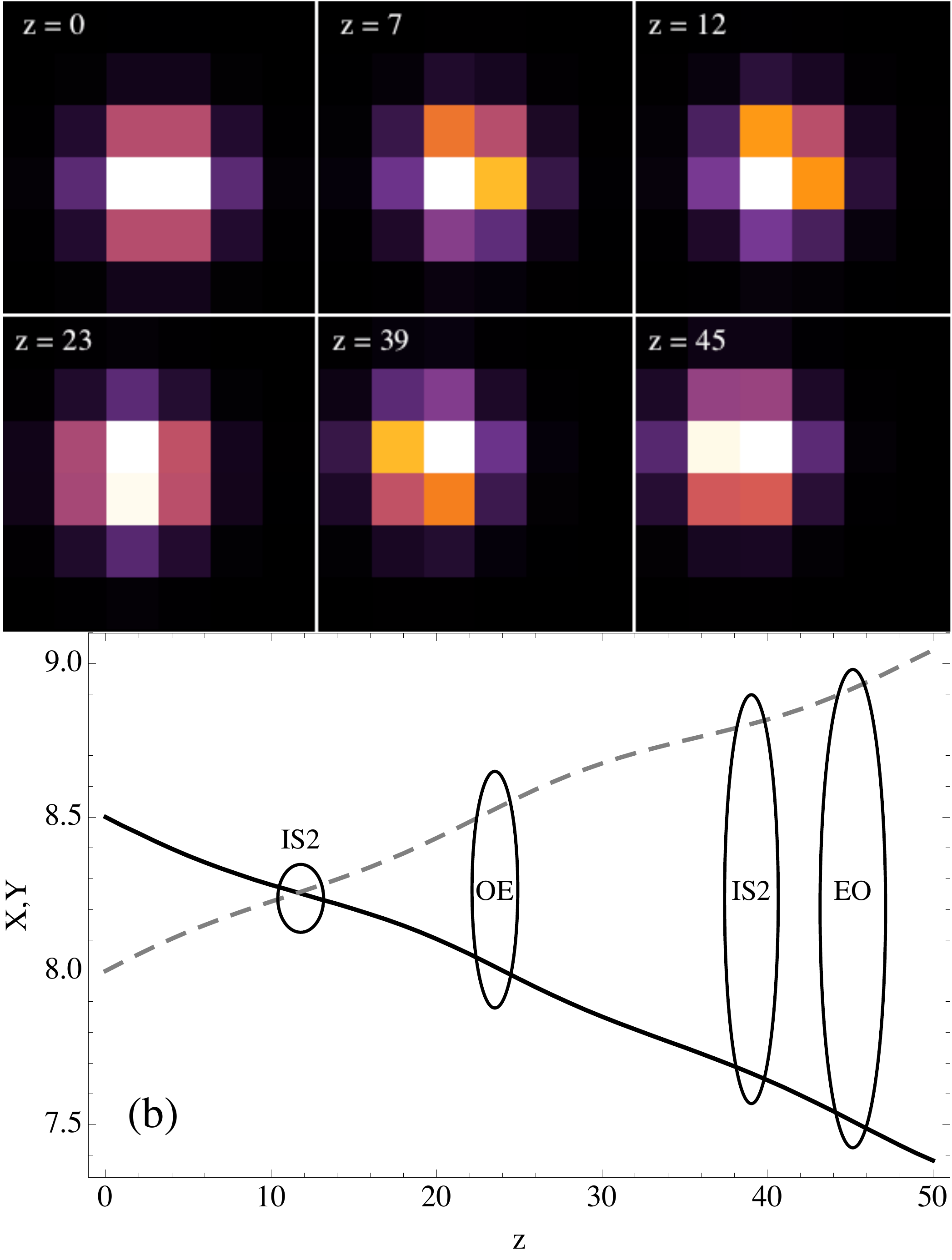}
\includegraphics[width=0.45\textwidth]{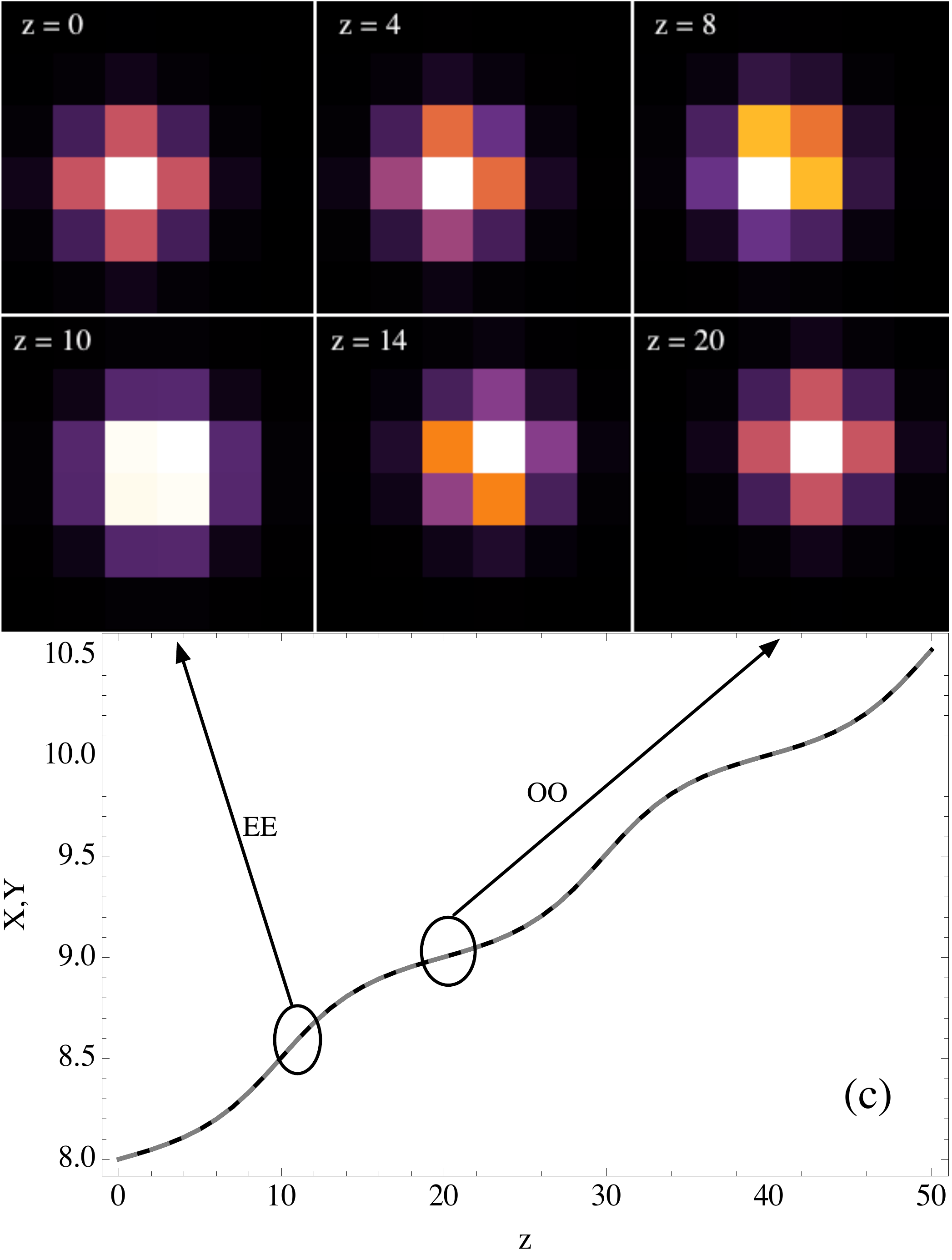}
\includegraphics[width=0.45\textwidth]{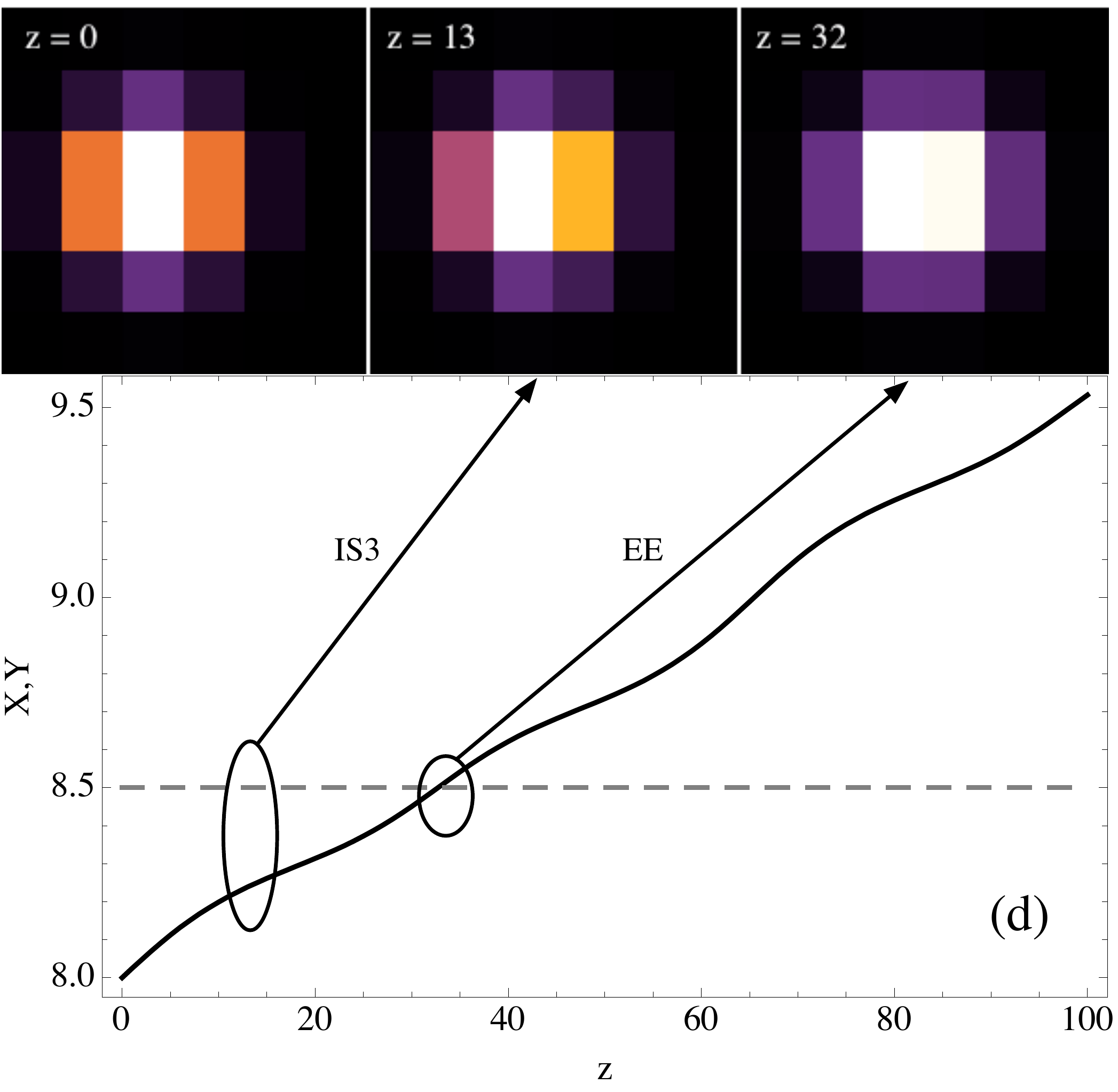}
\caption{(Color online) 
Examples of mobility dynamics in the propagation direction $z$, 
corresponding to surfaces shown in Fig.\ \ref{pot1} (a), (c), (d), and (e), 
respectively. In each subfigure, top figures show profiles movement [(a) shows a colormap where colors were normalized to the maximum amplitude of each plot; this colormap also applies to (b)-(d)], 
and bottom figures center of mass evolution for $X$ (full line) and $Y$ 
(dashed line). (a) $P=5$, $k_x=k_y=0.038$; (b) $P=10$, $-k_x=k_y=0.018$; 
(c) $P=12$, $k_x=k_y=0.015$; (d) $P=35.5$, $k_x=0.015$ and $k_y=0$. (Other 
parameter values are the same as in Fig.\ \ref{pot1}.)}
\label{dina1}
\end{figure*}
%
We first discuss an example from the first, low-power, regime 
corresponding to the energy surface in Fig.~\ref{pot1} (a) 
(as mentioned above, different mobility examples from this regime were 
shown in Figs.\ 4 (a) and (c) in Ref.\ \cite{pre73}). 
Here, we took the (unstable) EO solution and kicked it with a very small value 
in 
both directions, yielding the dynamics  
numerically observed in Fig.~\ref{dina1} (a). 
In the  $y$-direction, the kinetic energy is not sufficient to 
overcome the barrier created by
the four-site solution, while in the $x$-direction
it can move towards the minimum corresponding to the one-site solution.
The resulting dynamics for the center-of-mass positions shows 
how the movement in the horizontal direction gets combined with 
oscillations in the vertical one. 
The dependence $X$ vs $z$ also clearly shows how the 
solution feels the potential in terms of its velocity: maximum velocities 
occur around integer lattices sites (where the potential is a minimum) while 
the 
minimum velocities occur close to middle points 
[see full line in Fig.~\ref{dina1} (a); compare also with the analogous 
scenarios for the kicked OO mode in  Figs.\ 4 (a) and (c) of 
Ref.\ \cite{pre73}]. 

In the second power regime, with energy surfaces as in Fig.~\ref{pot1} (b), 
good mobility can be expected only in axial directions, with effective energy 
barrier determined by the intermediate solution between the one- and the 
two-site 
solutions as discussed in Ref.\ \cite{pre73}. The dynamics 
in this power regime, with a very small kick applied only in the axial 
direction to one of the stable stationary solutions,
will be as already illustrated in Fig.\ 4 (d) of Ref.\ \cite{pre73}:
the solution moves very slowly and adiabatically traces the shape of the 
potential with a minimal velocity at the places of the intermediate solutions.

In the third power regime, with surfaces as in Fig.~\ref{pot1} (c),
a very interesting kind of mobility is observed: a 
diagonal 
mobility between the (stable) horizontal and  vertical two-site solutions
as illustrated in  Fig.~\ref{dina1} (b). The initial
EO solution, kicked equally 
in the $(-x)$- and $y$-directions, 
gets sufficient kinetic energy to pass over the small barrier created 
by the intermediate IS2 solution. It then continues through the OE solution, 
passes another IS2 barrier, and then to the other EO 
solution shifted by one site in both directions. 
Although 
the potential connecting these two solutions is not completely flat, 
there is a very good 
transport of energy in this direction, allowing mobility for more than one 
lattice diagonal in the considered $15 \times 15$ lattice. 

If we take a look at the surface in the fifth power regime 
[Fig.~\ref{pot1} (d)], we  
realize that an initial (unstable) 
one-site solution may move in any direction by 
slightly kicking it since it corresponds to a maximum. 
Fig.~\ref{dina1} (c) shows an example for kicking the 
one-site solution symmetrically in both directions in order to make it move 
passing through the four-site, i.e., a diagonal movement. 
The velocity has maximum in the minima 
of the potential (corresponding to the EE solution) and minimum in the 
potential maxima (OO solution) (no intermediate solutions appear in this 
regime).

Finally, in the fourth power regime, 
we see from the surface [Fig.~\ref{pot1} (e)] that the 
two- and four-site solutions are stable simultaneously, presenting an 
intermediate solution in-between that will define the effective energy barrier. 
This barrier is very small and, therefore, a very small kick is also needed. 
Figure~\ref{dina1} (d) shows the evolution starting from a two-site vertical 
solution, passing through the intermediate one and arriving to the four-site 
solution.

As illustrated by the example in Fig.\ \ref{dina1}(a), the 
energy surfaces also provide nice intuitive interpretations to the dynamics 
of moving discrete solitons with additional perturbations 
{\em transverse} to the direction of motion. 
In this 
example, the dynamics in the two orthogonal directions 
appear essentially independent of each other 
(propagation in $X$ while oscillating in $Y$). 
However, as we will illustrate 
with another example below, there are situations where the particular 
surface topologies may lead to a more intricate interplay between the 
oscillatory and translational dynamics. 

\begin{figure}[htbp]
\centering
\hspace{0.58cm}
\includegraphics[width=0.47\textwidth]{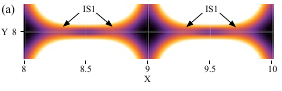}
\includegraphics[width=0.47\textwidth]{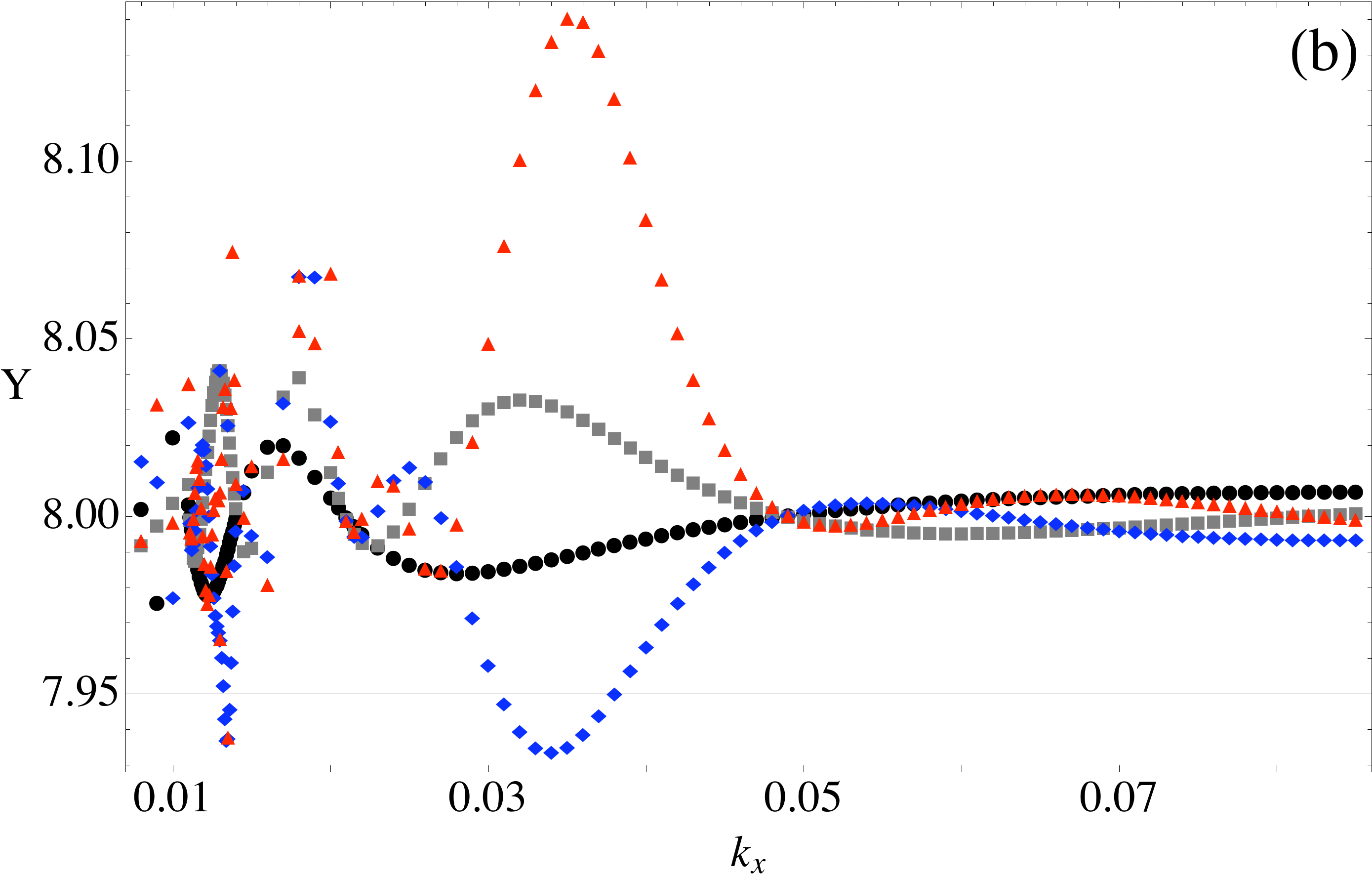}
\includegraphics[width=0.47\textwidth]{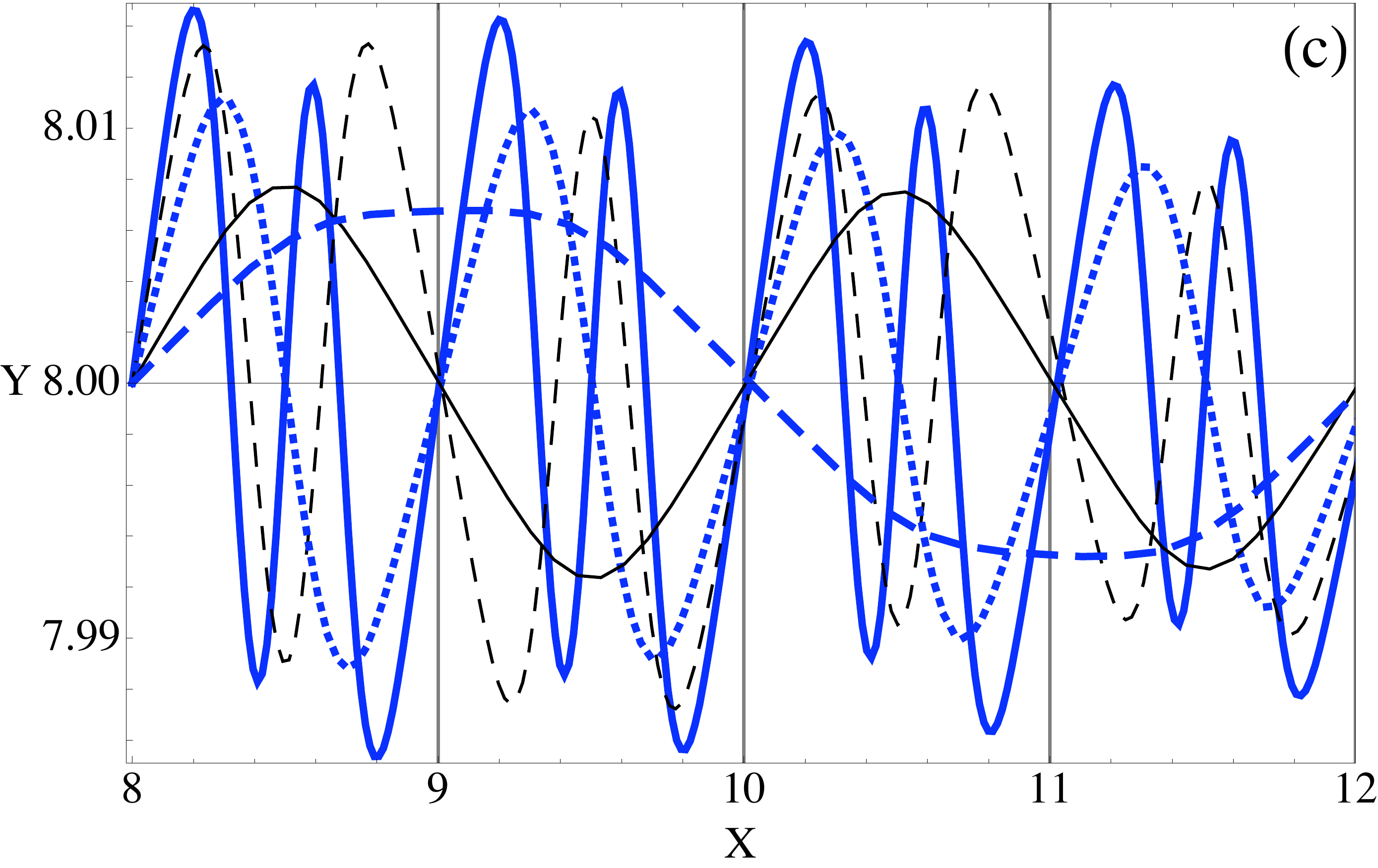}
\includegraphics[width=0.47\textwidth]{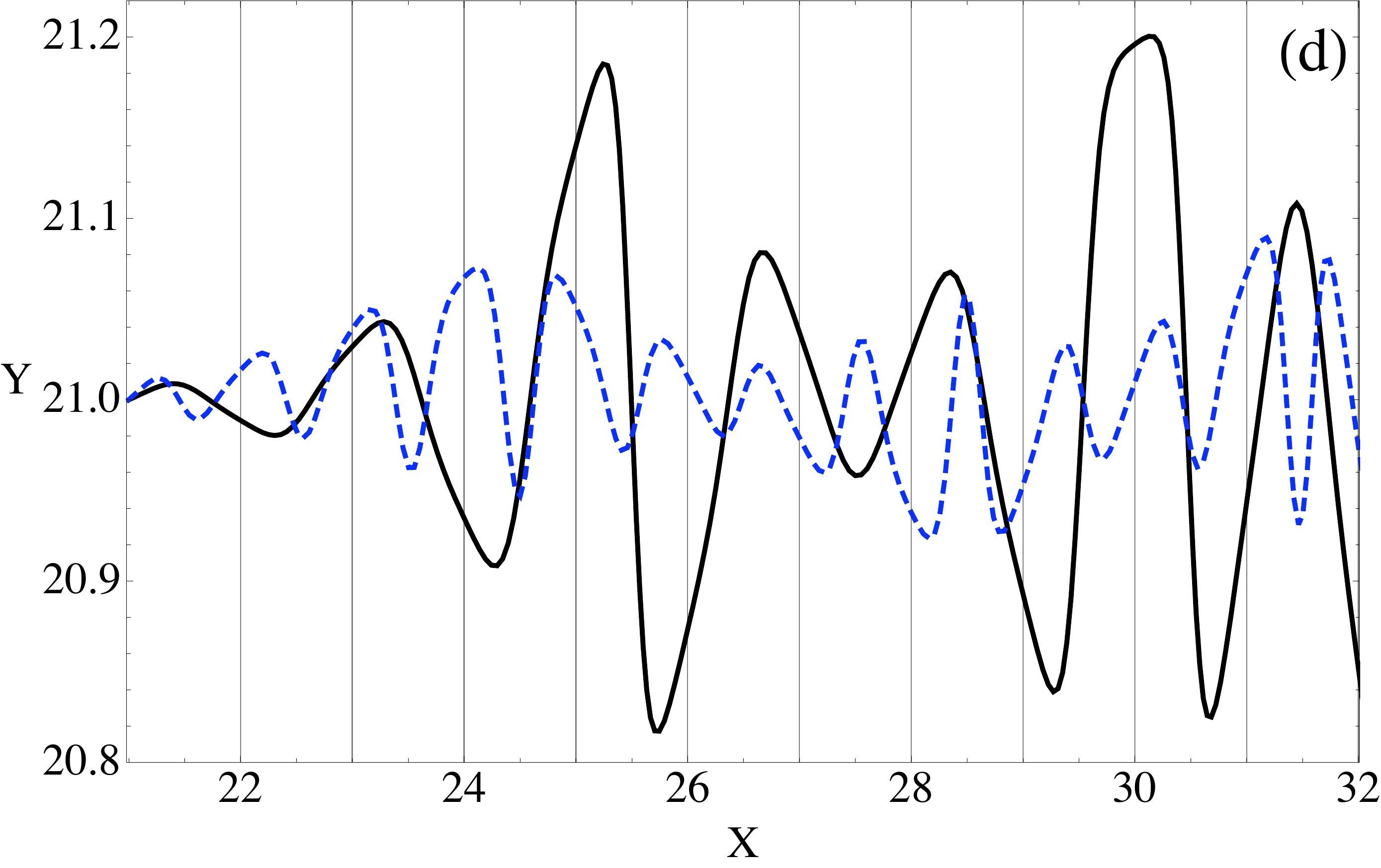}
\caption{(Color online)
Dynamics of an OO mode with $P=44$ ($\gamma=4, \alpha=1$), 
kicked with $k_y=0.001$ and varying $k_x$.
(a) Extended energy surface. Darker (brighter) regions 
correspond to lower (higher) values of the energy $H$. 
(b) $Y_9,\ Y_{10},\ Y_{11},\ Y_{12}$ versus $k_x$ represented by (black) 
filled circles, 
(grey) squares, (blue) diamonds and (red) triangles, respectively. 
(c) $Y(z)$ versus $X(z)$ for 
$k_x=0.0111,\ 0.0141,\ 0.021,\ 0.049$ and $0.083$ represented by 
thick (blue) solid, 
thin (black) dashed, (blue) dotted, thin (black) solid and 
thick (blue) dashed lines, respectively. 
(d) Dashed (blue) and solid (black) lines correspond to $k_x=0.018$ and $k_x=0.035$, 
respectively, for $N=M=41$.}
\label{uns}
\end{figure}

We present in 
Fig.\ \ref{uns} a case with $P=44$.  
With the notation from Sec.\ \ref{surfaces}, this value 
of the power belongs to the second power regime, and 
the structure of the surface is phenomenologically 
identical to the one 
sketched in Fig.\ \ref{pot1}(b), but for a higher level of power. 
A picture of this energy landscape, periodically extended along 
the $X$-direction, 
is shown in Fig.\ \ref{uns} (a). 
In this figure, the darkest regions 
correspond to the positions for the minima (the OO and EO/OE modes),  
while the brightest regions correspond to 
maxima (EE modes). Note also the positions of the saddle points 
(IS1, marked with arrows) in-between the OO and EO/OE modes. 

%

The kick we choose for the vertical direction is very small, 
$k_y=0.001$, in order to disturb the horizontal movement only weakly. 
Since the potential shape in the $Y$-direction - in a first approximation - will 
correspond to a harmonic potential, the profile will essentially 
perform harmonic oscillations in the $Y$-direction, with a frequency (determined by the curvature of the surface) 
that will vary only slightly as the mode 
translates in the $X$-direction. 
To look for possible resonances between the oscillatory and translational 
dynamics,
we implement the following scheme: 
For fixed $k_x$ and 
$k_y$ we numerically integrate model (\ref{sdnls}), starting with the OO 
stationary solution centered at position $\{8,8\}$, and measure the z-values for 
which $X$ is equal to $9, 10, 11$ and $12$, respectively. 
For these z-values we compute 
$Y_m\equiv Y(X=m)$, and plot in Fig.\ \ref{uns}(b) as a function of the 
horizontal kick strength $k_x$.

The particular values of $k_x$ where symbols in Fig.\ \ref{uns}(b) 
coincide for $Y=8$ correspond 
to kick strengths where the natural frequency of the surface potential
is commensurate with the translational velocity.
In Fig.\ \ref{uns}(c) we show 
some examples for these particular points. 
For $k_x=0.0111$ the solution makes two cycles 
before arriving to the next integer position in $X$. For $k_x=0.0141$ the 
solution makes one and a half cycles, for $k_x=0.021$ one cycle, for 
$k_x=0.049$ one half of cycle and for $k_x=0.083$ one quarter of cycle, before 
arriving to the next integer position in $X$. As can be seen, around these 
points the oscillations apparently remain bounded and the dynamics stable, 
at least for long times. (Note that we do not see any visible phase-locking 
effects between the translational and oscillatory motion here: the symbols in 
Fig.\ \ref{uns}(b) seem to coincide only at isolated points and not in 
intervals. This may be an indication that the dissipative effects on the 
effective dynamics are very weak.)

However, in other regimes we observe that the amplitude of the oscillations 
in the $Y$-direction rapidly grows. Two examples are illustrated in 
Fig.\ \ref{uns}(d), corresponding to values of  $k_x$ 
where the $Y$-displacement in Fig.\ \ref{uns}(b) has maxima. 
(Here, we have used a larger system  of $41\times 41$ sites, 
in order to clearly identify the nature of this particular dynamics.) 
The initial increase of the $Y$-amplitude suggests a kind 
of unstable dynamics. However, as is clear from the longer simulation 
in  Fig.\ \ref{uns}(d), the oscillation amplitude in the Y-direction remains
bounded and thus the trajectory does not escape in the vertical direction but 
remains trapped by the barrier.
The typical dynamics, thus, exhibits amplitude-modulated oscillations 
(beatings), with the largest amplitudes corresponding to maxima close to 
integer $X$ where the surfaces are flatter. 
Thus, this kind of mobility is a signature of the 2D topology of the 
potential and it also validates the surfaces computed with our method.
(We performed several computations in different $\{P,k_x\}$-regimes finding 
similar results as the ones presented in Fig.\ \ref{uns}).

\section{Effects of weak anisotropy}\label{anisotropy}

\begin{figure}[htbp]
\centering
\includegraphics[width=0.48\textwidth]{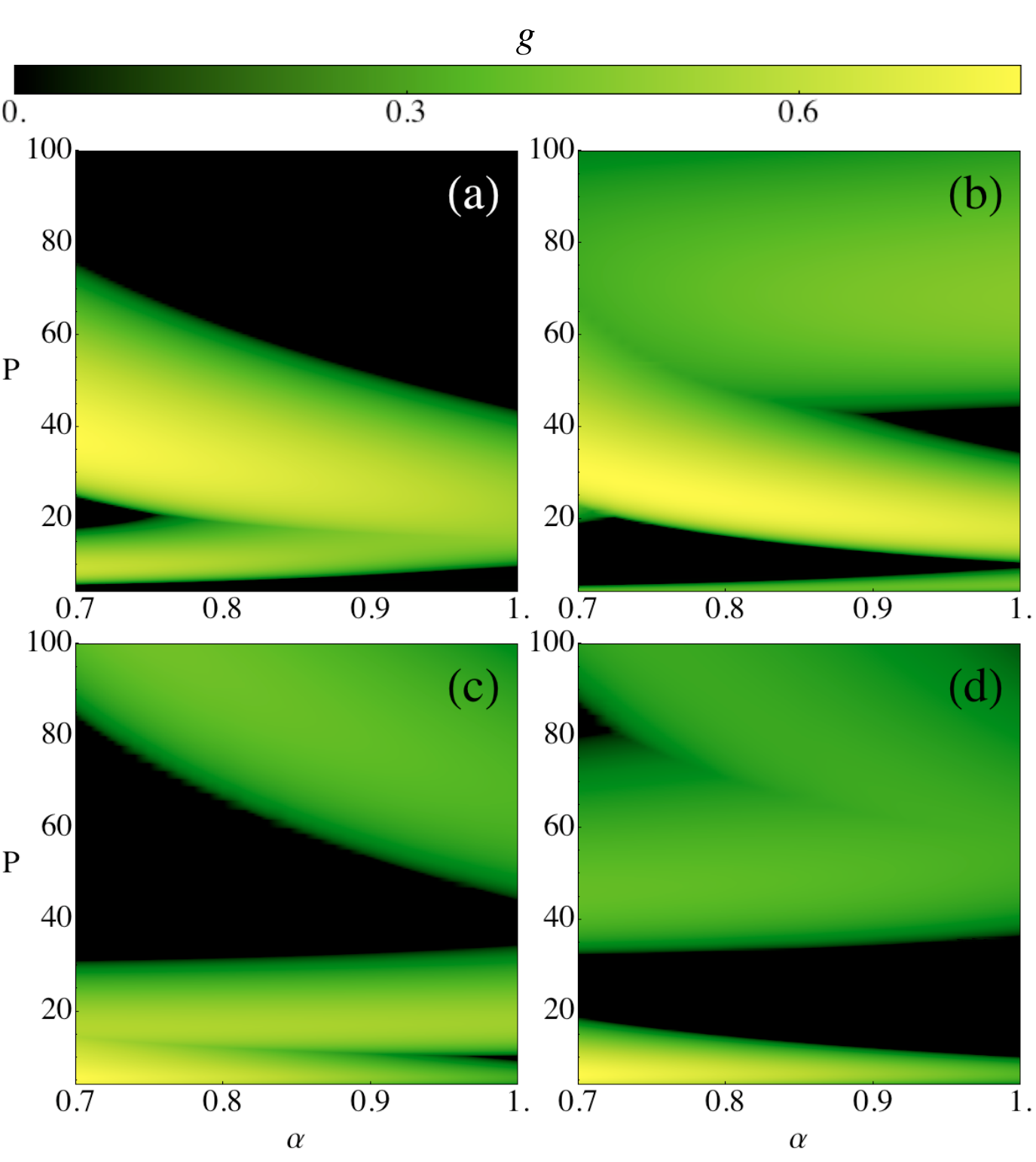}
\caption{(Color online) 
Density plots of $g$ versus $P$ and $\alpha$ for $\gamma =4$, for
the (a) OO; (b) EO; (c) OE; and (d) EE solutions, respectively. 
Black means $g=0$ and lighter colors imply an increasing instability.}
\label{stabi}
\end{figure}
%
Considering weak anisotropy by putting 
$\alpha < 1$ in Eq.\ (\ref{sdnls})
(implying a stronger coupling in the horizontal than in the vertical 
direction), 
the main qualitative modifications as concerns the 
stationary solutions are shifts of the 
stability regions for the different solution types (evidently, the
two-site horizontal, EO, and vertical, OE, solutions are now non-equivalent).
Focusing our discussion on the parameter value  $\gamma =4$ analyzed 
in detail for the isotropic case in the previous sections, 
the results for the stability analysis are presented in Fig.\ \ref{stabi}.
We find that, for fixed $\alpha < 1$ and increasing power,  
the first stability inversion appears between the one-site OO
and the horizontal two-site EO solutions, and it occurs for lower values 
of the power than in the isotropic case. 
The first interval of stability of the four-site EE 
solution gets narrower when $\alpha$ decreases, mainly because its lower limit 
increases (with 
a corresponding increase of the upper stability limit for the EO solution). 
The first 
stability region of the vertical two-site OE solution disappears for
$\alpha \lesssim 0.96$. On the other hand, its second stability regime for 
higher powers 
is enhanced, whereas the corresponding stability regime for its horizontal 
counterpart gets narrower and disappears for $\alpha \lesssim 0.89$.

\begin{figure}
\includegraphics[width=0.45\textwidth]{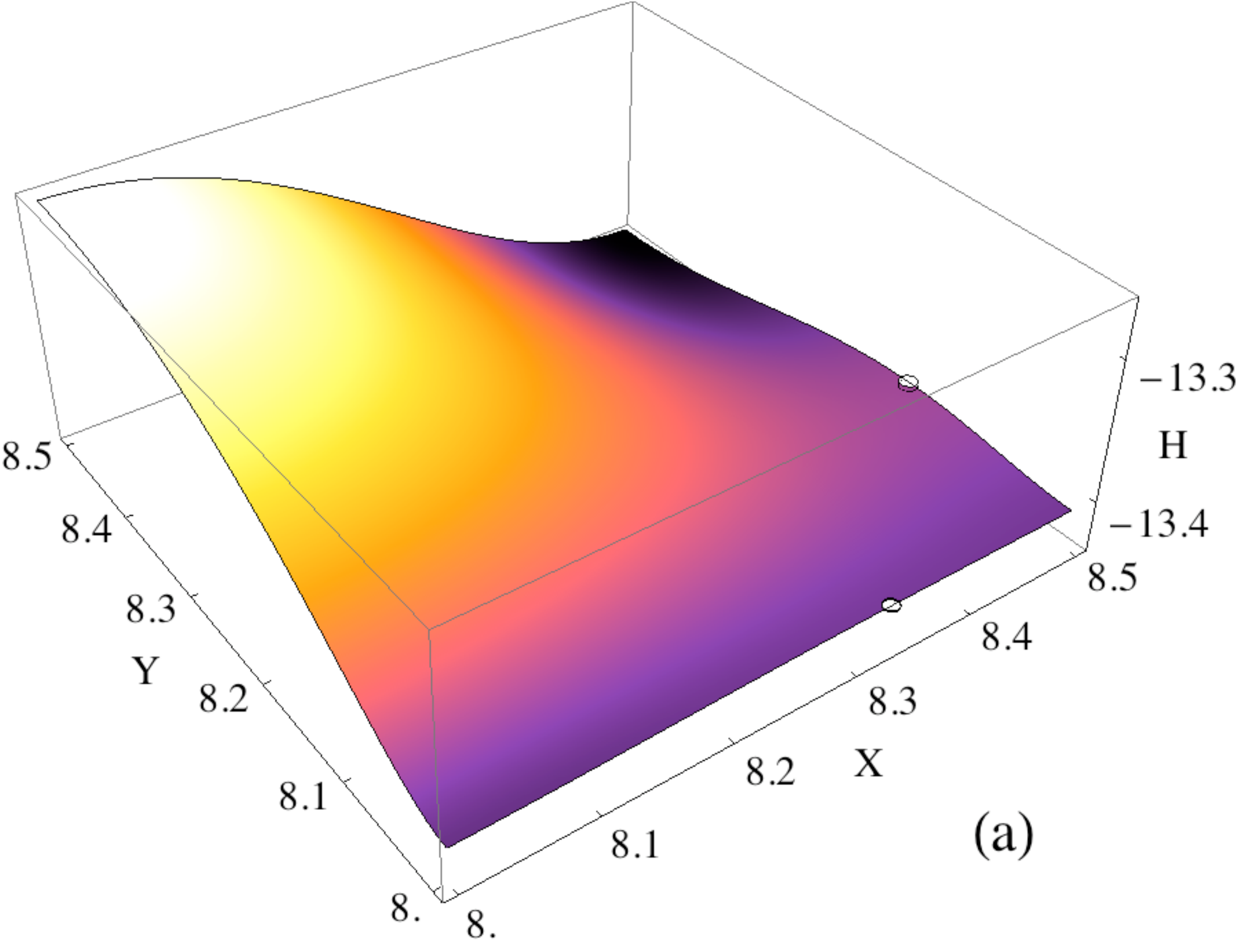}
\includegraphics[width=0.45\textwidth]{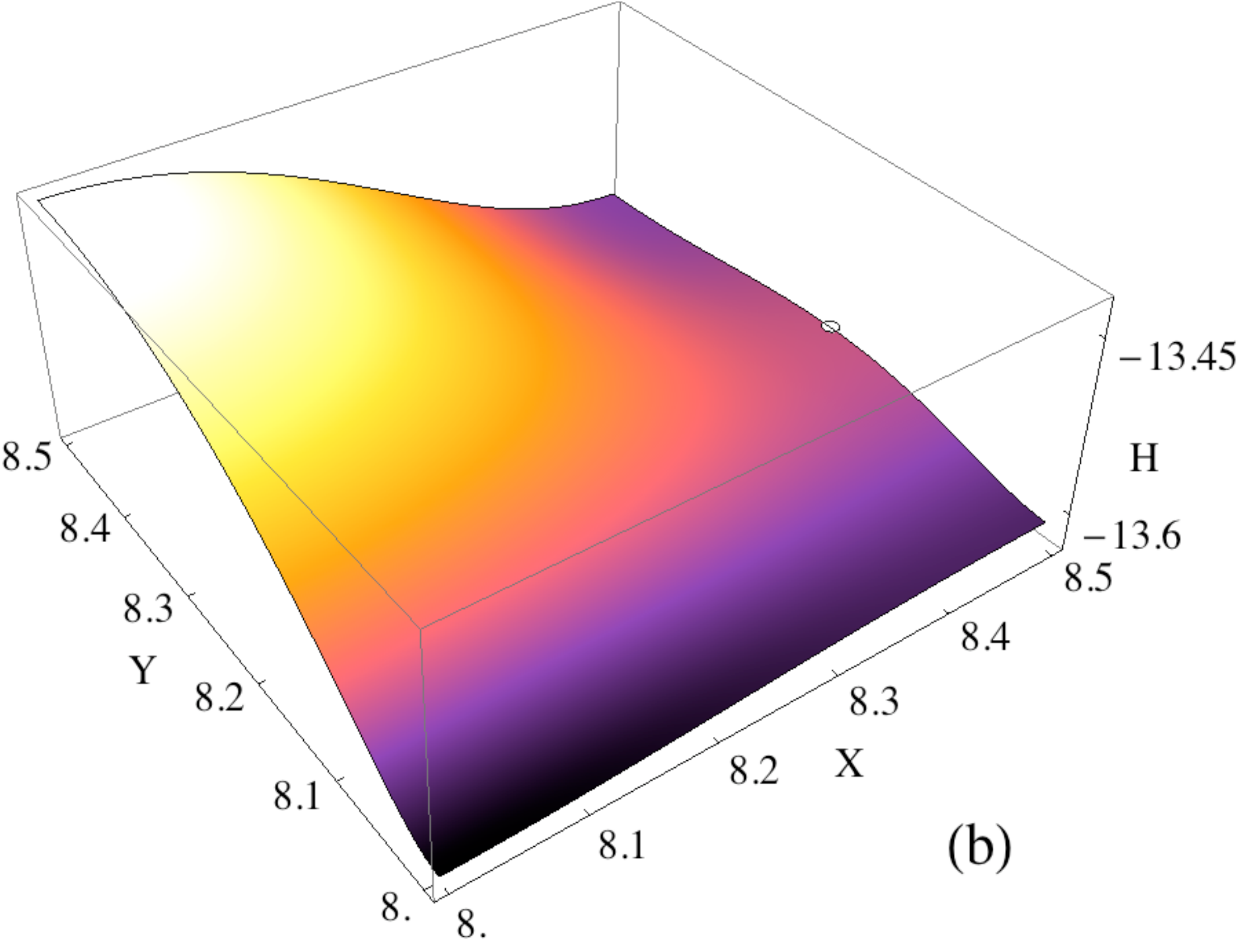}
\caption{(Color online) 
Energy surfaces for anisotropic lattices with $\gamma=4$: 
(a) $\alpha=0.72$, $P=19.5$; (b) $\alpha=0.7$, $P=20$.}
\label{anpot1}
\end{figure}
With stronger anisotropy, some qualitatively new regimes not present in the 
isotropic case will appear. For $\alpha \lesssim 0.76$, a new region of 
stability of the OO mode appears around  $P\approx 20$, 
leading to a small window of multistability where the 
OO, EO and EE modes are stable simultaneously. An example of an energy 
surface from this regime  
is shown in Fig.~\ref{anpot1} (a),  
where also two intermediate solutions [white dots] can be 
found on the 
edges. There is also a regime with  bistability of 
the OO and EE modes only, as illustrated by 
the pseudo-potential surface in Fig. \ref{anpot1} (b). 
Note that, in this case, the intermediate 
solution does not leave the edge, in contrast to
the IS2 mode existing on the diagonal between 
the one- and four- site solutions for the isotropic case of simultaneously 
stable EO/OE solution [Fig. \ref{pot1} (c)].
Since anisotropy destroys the symmetry between the OE and EO solutions, 
bistability of both 2-site solutions 
disappears already for a small amount of anisotropy, and for smaller 
$\alpha$ the IS position can only 
be found on an edge of the potential surface.

Thus, since no additional ``valleys'' were found to develop in the PN surfaces 
(all minima appear on the edges), the best mobility for anisotropic lattices 
should generally be expected
to occur along lattice directions (at least in cases where diagonal 
coupling terms can be neglected, as assumed throughout this work).

\section{Conclusions}\label{conclusions}

In conclusion, we have studied in a deeper way the problem of 
mobility of localized modes 
in two-dimensional saturable discrete systems. 
We numerically implemented
a constrained Newton-Raphson method 
to construct full Peierls-Nabarro energy 
surfaces, which appeared as very useful tools for predicting the dynamical 
properties of localized excitations. Although these surfaces were never found 
to be completely flat
(and therefore the corresponding Peierls-Nabarro barriers is strictly never zero), 
parameter regimes and directions of good mobility were seen to correspond 
to smooth, flat parts of the surfaces. 

For the isotropic saturable model, five different surface topologies 
could be identified in different power regimes, 
depending on the 
stability properties of the different fundamental stationary solitons. 
By numerically studying the dynamics of perturbed stationary 
solutions, we showed how these different topologies yielded qualitatively 
different kinds of optimal mobility,  
generally in axial or diagonal directions and with velocities varying 
according to the shape of the potential
while the profile is propagating across the lattice.  
The energy surfaces were also found to be very useful for interpreting the 
dynamics resulting from the interplay between translational and oscillatory 
motion in orthogonal directions. 

We also studied the effect 
of weak anisotropy, where the breaking of the symmetry of the system 
also yields energy surfaces with lower symmetry. In addition, we 
showed how this symmetry breaking could lead to a 
change of stability properties for the fundamental 
stationary solutions, and as a result two new types of surface topologies 
were identified. From the shape of these surfaces, we predicted that the
best mobility in the anisotropic case should generally occur along 
lattice directions.  

Thus, the method developed here appears to be a very good tool for the 
understanding of the mobility dynamics of localized 
excitations in higher-dimensional lattices, 
clearly showing the ``real'' energy barriers that localized solutions 
experience. 
We hope that this approach
will be useful also for other types of nonlinar lattice models. 

Finally, from an application point of view, we have given further examples 
on how the saturable nature of nonlinearity promotes the 
generation of many different stability and mobility scenarios. The example 
presented in section \ref{dynamics} shows the possibility to steer the 
dynamics of the solution by using 
the natural frequencies of its own self-induced potential. As an application 
of this, we propose a discriminative optical switch which, for 
example, will only be activated when $X$ and $Y$ positions are - 
simultaneously -
an integer number. This is one more example of nonlinear lattices being the 
key to success for controlling the propagation of light in future all-optical 
technologies.

\begin{acknowledgments}

We are grateful to Serge Aubry and Sergej Flach who both, independently, suggested this approach to analyze 2D mobility, and also for discussions and comments on the manuscript. We thank Mario I. Molina for useful discussions. The authors acknowledge financial support from different sources: FONDECYT 1070897 and 7080001, FB0824/2008, and CONICYT fellowship. Authors acknowledge the hospitality of the Max Planck Institute for the Physics of Complex Systems (MPIPKS), Dresden, where this work was in part developed. M.J. also acknowledges support from the Swedish Research Council. 
\end{acknowledgments}

\end{document}